\documentclass[fleqn, usenatbib]{mnras}
\usepackage[T1]{fontenc}
\usepackage{booktabs}
\usepackage{pgfplotstable}
\pgfplotsset{compat=1.15}
\pgfplotstableset{precision=4}
\usepackage{graphicx}
\usepackage{amsmath}
\usepackage{amssymb}
\usepackage{subcaption}
\usepackage{pdflscape}
\usepackage{xcolor}
\usepackage{anyfontsize}
\usepackage{hyperref}

\title[Electron Cyclotron Maser Emission in LPRTs]{Electron Cyclotron Maser Emission as the Driving Mechanism in Long-Period Radio Transients}

\author[L. Ferrario]{Lilia Ferrario$^{1}$\thanks{E-mail: lilia.ferrario@anu.edu.au} \\
$^{1}$Mathematical Sciences Institute, The Australian National University, Canberra, ACT 2601, Australia}

\date{Accepted 2026 January 15. Received 2026 January 15; in original form 2025 November 18}
\pubyear{2026}

\begin{document}
\maketitle
\begin{abstract}
Long-period radio transients (LPRTs) are highly polarised, coherent radio sources with periods of minutes to hours and bursts typically lasting $10-100$\,s. Here we consider the apparently isolated subclass of LPRTs and argue that electron cyclotron maser emission (ECME) explains their narrow duty cycles and polarisation properties. In particular, we show that intrinsically circular ECME can emerge as predominantly linear after undergoing Faraday conversion in an overlying magnetospheric plasma layer, thus reconciling the observed high linear fractions with a circularly polarised maser.  In this picture, a rotating oblique magnetosphere beams radiation into a thin, hollow emission cone whose surface lies almost perpendicular to the local magnetic field. The observed very narrow pulses arise when the line of sight skims the cone, while broader profiles and weak leading or trailing components occur when multiple azimuths along the emission ring meet the maser resonance condition. The observed isotropic-equivalent luminosities of $\sim10^{30}-10^{31}$\,erg\,s$^{-1}$ correspond to modest intrinsic powers once strong ECME beaming is taken into account. We show that such power levels can be supplied by accretion from the interstellar medium (ISM), and that detectability at kiloparsec distances favours slowly rotating neutron stars with comparatively low surface magnetic fields ($B_\ast\lesssim10^{10}$\,G) and low space velocities.
\end{abstract}


\section{Introduction}

LPRTs are a heterogeneous class of radio sources characterised by highly polarised, coherent bursts with repetition timescales of minutes to hours. In the literature, this phenomenological label encompasses a variety of systems, including interacting binaries such as the magnetic cataclysmic variables (MCVs) and white-dwarf pulsars, as well as apparently isolated objects. In this work, we focus specifically on the subclass of LPRTs that are \emph{apparently isolated}, exhibit stable long periods, narrow radio duty cycles, and extreme polarisation fractions. Their periodicities and polarisation properties distinguish them from ordinary pulsars and magnetars, strongly suggesting a different emission mechanism. Moreover, their inferred brightness temperatures, $\gtrsim10^{16}$\,K, require a coherent emission process. They are typically detected at frequencies of $\sim0.2-1.4$\,GHz, with some sources observed up to $\sim2$\,GHz. Peak flux densities are in the range $0.1-40$\,Jy, implying isotropic-equivalent radio luminosities of $\sim10^{30}-10^{31}$\,erg\,s$^{-1}$ for typical distances of $\sim1-5$\,kpc \citep{ReaHurley-Walker2024_LPRPs_Review}. 

The prototypical source, GLEAM-X\,J$1627-5235$ \citep{Hurley-Walker2022_GLEAM-X1627}, shows strongly linearly polarised (up to nearly $90$ per cent) bursts lasting $30-60$\,s every $18.18$\,min. Subsequent discoveries include GPM\,J$1839-10$ ($P=21.97$\,min; \citealt{Hurley-WalkerRea2023_LPRT}), ASKAP\,J$1755-2527$ ($P\sim 1.16$\,h; \citealt{Dobie2024_ASKAPJ175534-252749,McSweeney2025_ASKAPJ175534-252749}), ASKAP\,J$1935+2148$ ($P=53.8$\,min; \citealt{Caleb2024_LPRT_54min_Switching}), and ASKAP\,J$1839-0756$ ($P=6.45$\,h; \citealt{Lee2025_6.45hours_LPRT_Interpulses}), the latter showing a weaker interpulse separated by half a rotation. A further source, ASKAP/DART\,J$1832-0911$ emits minute-long radio bursts (occasionally extending to several minutes), every $44$\,min that are accompanied by phase-aligned X-ray pulses \citep{Li2024_DARTJ1832-091_SNR,WangRea2025_ASKAPDartJ1832-0911DetectionX-rays} .

Despite differences in period and polarisation characteristics, LPRTs share several features: (i) extremely high brightness temperatures, (ii) lack of detected emission at other wavelengths (with the exception of ASKAP/DART\,J$1832-0911$), (iii) high and variable polarisation, and (iv) long-term phase stability. These properties make standard pulsar or magnetar interpretations rather difficult. Canonical dipole spin-down, under standard assumptions, cannot produce neutron stars with spin periods of minutes to hours within the age of the Galaxy, unless the neutron star possesses ultra-strong magnetic fields ($B_\ast\gtrsim10^{15}$\,G) that do not decay significantly over time. Moreover, even if such long spin periods are attained, rotation-powered activity becomes intrinsically inefficient. Standard pulsar-like rotation-powered particle acceleration is ineffective at the long periods of LPRTs: the available potential drop scales as $P^{-2}$ and becomes too small to sustain pair cascades or maintain the relativistic particle flux required for coherent emission. Likewise, isolated MWDs, when treated as rotation-powered systems, generate voltages that are orders of magnitude below those needed to lift and accelerate charges from the stellar surface. In both cases, the available magnetospheric particle supply falls below that required to sustain coherent radio emission, rendering rotation-powered mechanisms ineffective at the observed luminosities and polarisation levels \citep[see][]{ReaHurley-Walker2024_LPRPs_Review}. 

Given the narrow duty cycles of LPRTs, stable periodicities, and high polarisation, the ECME mechanism could be the most convincing explanation for their coherent radio emission. Several physically distinct pathways can lead to ECME, and it is important not to conflate them. First, ECME is well established in planetary (including terrestrial) and stellar magnetospheres, where it is powered by magnetospheric currents and driven by loss-cone or horseshoe anisotropies in the electron distribution \citep[e.g.][]{Zarka1998_AuroralEmission_OuterPlanets, Leto2016_AuroralRadioEmission, Trigilio2011_CuVir_maser}. 

Second, ECME has been invoked in compact interacting binaries hosting a MWD. In MCVs \citep{FerrarioWickKawka2020_MWDs}, coherent radio bursts are attributed to ECME, although the precise emission site remains debated \citep{Barrett2020AdSpR_RadioObs_MCVs_Maser}. In binary systems such as AR Scorpii \citep{Marsh2016_ARSco_Discovery}, ECME may operate in a pulsar-like interaction between a rapidly spinning MWD and its low-mass companion. A further, physically distinct, binary-driven scenario proposes unipolar induction in non-interacting WD-M dwarf pairs, in which the electromotive force generated along magnetic loops linking the two stars powers ECME \citep[e.g.][]{Horvath_2025_ECME_Unipolar_MCVs}. This has been suggested for long-period radio sources in confirmed compact binaries, such as ILT\,J$1101+5521$ \citep{Qu2025__ILT1101_unipolarInductor}. These scenarios, however, rely on binarity and are therefore not applicable to the isolated LPRTs considered here.

The present paper examines a different regime entirely, namely ECME from \emph{isolated} neutron stars and, more speculatively, isolated MWDs. In this picture, the maser is fuelled by a continuous supply of mildly relativistic electrons provided by accretion from the ISM. This scenario naturally accounts for the lack of companions in many LPRTs, and for the presence of soft X-ray emission in at least one source.

In isolated LPRTs, ECME likely originates in the outer magnetospheres of neutron stars or MWDs where the local magnetic field satisfies
\[
\nu_{\rm ce}\approx2.8\,\mathrm{MHz}\left(\frac{B}{1\,\mathrm{G}}\right),
\]
as first derived by \citet{WuLee1979_EarthKmRadiation} in the context of terrestrial auroral kilometric radiation. Hence, GHz emission corresponds to $B\sim10^{2}-10^{3}$\,G. Such field strengths occur at radii of tens to hundreds of stellar radii, depending on the surface magnetic field of the star, with neutron stars typically having surface fields of $\sim10^8-10^{15}$\,G \citep[e.g.,][]{HardingLai2006_Review_PhysicsStronglyMagneticNeutronStars} and high-field MWDs $\sim10^6-10^9$\,G \citep[e.g.,][]{FerrarioWickKawka2020_MWDs}.

The possibility of accretion from the ISM was originally proposed by \citet{Ostriker1970_OldNeutronStars_ISM} as a way to detect extinct pulsars via their soft X-ray emission. The detection of X-rays from ASKAP\,J$1832-0911$ is consistent with weak, magnetically channelled accretion from the ISM that simultaneously powers the radio bursts and the high-energy emission.

In what follows, we present an ECME interpretation for LPRTs. Section~\ref{sec:ECME} summarises the maser mechanism and its geometrical implications. Section~\ref{sec:polarisation} discusses the origin of their polarisation diversity. Section~\ref{sec:longP_NS_ISM} discusses how neutron stars can reach the long spin periods of LPRTs and the possible power source. We discuss our results in Section \ref{sec:discussion}, while Section \ref{sec:conclusions} summarises the paper.

\section{Electron Cyclotron Maser Emission}\label{sec:ECME}

Here we summarise the geometry, while background information on ECME is provided in Appendix~\ref{app:ecme_essentials}.

This maser mechanism operates under the following resonance condition \citep{WuLee1979_EarthKmRadiation}:
\begin{equation}
\omega - k_{\parallel}v_{\parallel} - \frac{n\,\Omega_{e}}{\gamma} = 0,
\end{equation}
where $\omega$ is the wave angular frequency, $k_{\parallel}$ is the component of the wavevector $\mathbf{k}$ parallel to the local magnetic field $\mathbf B_{\rm loc}$, $v_{\parallel}$ is the component of the particle's velocity $\mathbf{v}$ parallel to $\mathbf B_{\rm loc}$, $\Omega_{e}=eB_{\rm loc}/(m_{e}c)$ is the electron gyrofrequency, $\gamma$ is the Lorentz factor, and $n=1,2,3,\dots$ is the cyclotron harmonic number.

Maximum growth occurs for quasi-transverse propagation, that is, the emission peaks at $\theta_{Bk}\simeq 90^{\circ}-\delta\theta$, where $\theta_{Bk}$ is the angle between $\mathbf{k}$ and $\mathbf{B}$. The small offset $\delta\theta$ (typically a few degrees) is primarily set by the intrinsic ECME growth physics: mildly relativistic electrons with an unstable distribution emit in a hollow cone whose aperture deviates slightly from perfect orthogonality \citep{Hess2008Io-Jupiter_decameter-arcs}. Additional propagation effects, such as magnetospheric refraction, can further modify the apparent beaming direction and introduce frequency-dependent pulse arrival times \citep{Trigilio2011_CuVir_maser}. The finite wall thickness, $\sigma$, of the hollow cone, typically $\lesssim 1^\circ$, traces the angular width over which the growth rate remains significant.

Efficient maser amplification requires a low plasma density such that $\omega_{p}/\Omega_{e}\lesssim0.1-0.3$, where $\omega_{p}=(4\pi n_{e}e^{2}/m_{e})^{1/2}$ is the plasma frequency, with the upper limit depending on the electron velocity distribution function and harmonic. In this regime, the fundamental harmonic ($n=1$) generally dominates because it maximises the growth rate and minimises gyroresonant absorption. However, the second harmonic ($n=2$) can remain viable when the plasma density is moderately higher, the loss-cone anisotropy is strong, or propagation effects suppress fundamental escape, as is commonly observed in planetary and stellar ECME sources \citep{Treumann2006_ECME}.

\begin{figure}
    \centering
    \includegraphics[width=0.95\linewidth]{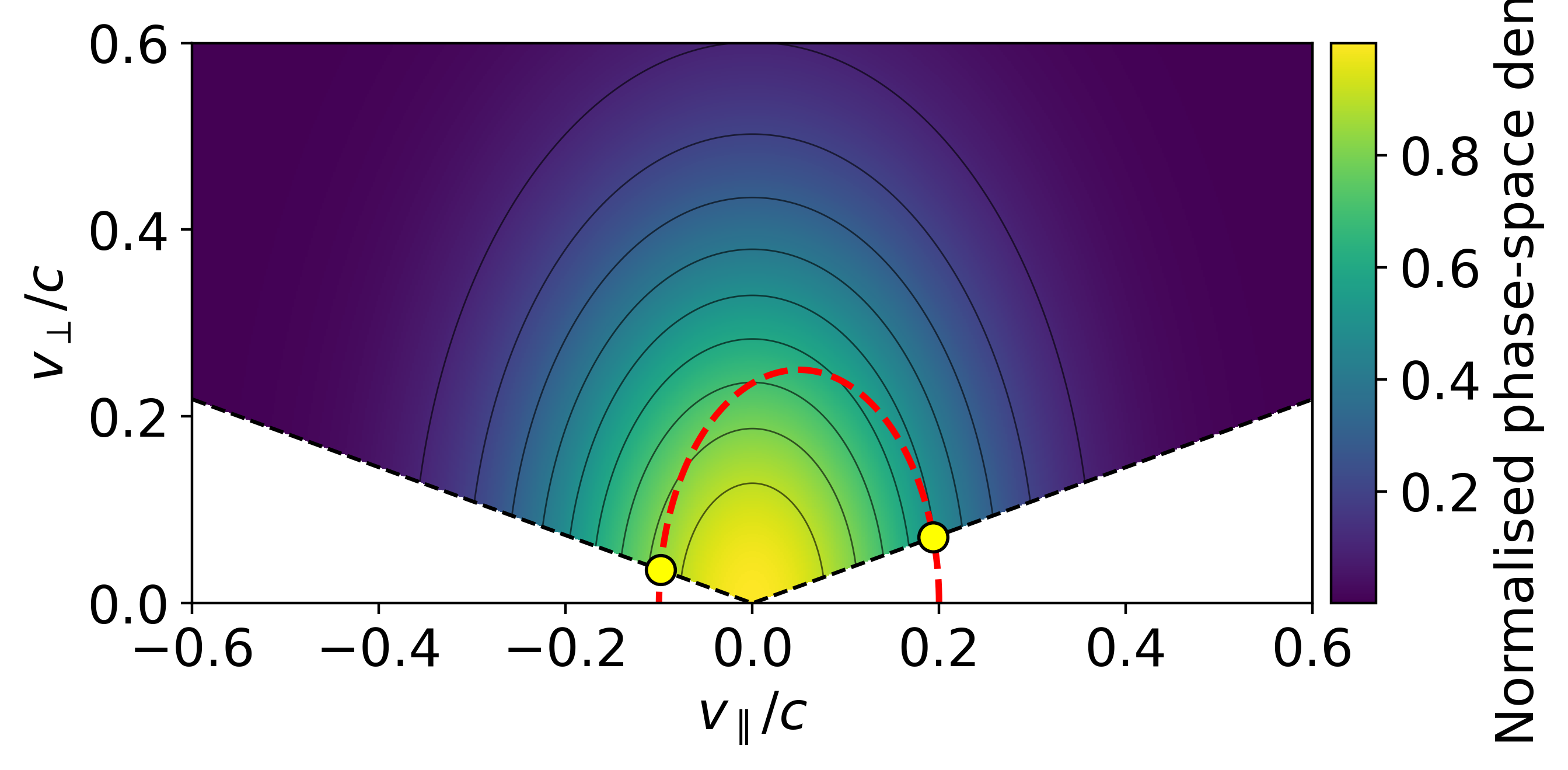}
    \caption{Illustration of the loss cone anisotropy and resonance ellipse in velocity space. Colour shading shows a bi-Maxwellian with symmetric loss cones (interior excised). Dashed lines mark the loss-cone edge $v_\perp= |v_\parallel| \tan\alpha_{\rm L}$. The red curve is a representative cyclotron resonance ellipse for fixed $(\omega,k_\parallel)$ and harmonic $n$. Yellow circles indicate the intersections where the resonance reaches the largest positive $\partial f/\partial v_\perp$, maximising maser growth at quasi-transverse propagation ($k_\parallel\simeq0$). }
    \label{fig:losscone-resonance}
\end{figure}

The so-called ``loss cone'' is the deficit of electrons at small pitch angles produced by magnetic mirroring (see Appendix \ref{app:ecme_essentials}). Particles with insufficient perpendicular momentum stream along the field lines toward regions of stronger magnetic field and are removed from the maser-active region, leaving a depletion in the distribution $f(v_{\parallel},v_{\perp})$ at low $v_{\perp}$. This creates the positive perpendicular gradient required for maser growth. Consequently, the growth rate in velocity space is proportional to the gradient $\partial f/\partial v_{\perp}$ of the electron distribution, so the maser amplifies radiation most efficiently where the resonance curve samples regions of positive perpendicular gradient. This is illustrated in Fig.~\ref{fig:losscone-resonance}, which summarises how the cyclotron resonance ellipse intersects the loss-cone boundary near $v_{\parallel}\simeq0$, selecting quasi-transverse propagation and thereby maximising the maser gain. Although the background shading in Fig.~\ref{fig:losscone-resonance} shows a bi-Maxwellian distribution, the existence and geometry of the loss cone do not depend on this choice and would arise for any reasonable parent distribution (e.g., a power law). We adopt the convention that $v_{\parallel}>0$ corresponds to motion away from the mirror point in one magnetic hemisphere, whereas $v_{\parallel}<0$ represents the analogous population in the opposite hemisphere.

The characteristic field at the emission site (see Appendix~\ref{app:ecme_essentials}) is
\begin{equation}\label{eq:Bloc}
B_{\rm loc}\,[\mathrm{G}]\simeq357\,\frac{\gamma}{n}\,\nu_{\rm obs}\,[\mathrm{GHz}].
\end{equation}
Thus, for $\nu_{\rm obs}$ in the range $1.0-1.4$\,GHz, $n=1-2$, and $\gamma\simeq1.1-2$, the inferred field strengths ($B_{\rm loc} \sim10^{2}-10^{3}$\,G) place the GHz ECME layer at radii of tens to hundreds of stellar radii in both neutron stars and MWDs, since for a dipole
\begin{equation}
\frac{r}{R_\ast}=\left(\frac{B_\ast}{B_{\rm loc}}\right)^{1/3}.
\end{equation}
As the star rotates the emission cone, whose axis is aligned with $\mathbf{B}_{\rm loc}$ at the emission site, sweeps across the observer’s line of sight. Pulses occur when the line of sight intersects the cone wall (see  Section~\ref{sec:geometry}).

\subsection{Geometry and visibility}\label{sec:geometry}

ECME is detected only when the beaming angle $\theta_{Bk}$ lies close to the characteristic cone angle $\theta_{\rm cone}$, where maser growth is strongest, so visibility depends sensitively on the observer’s line of sight and on the magnetic obliquity $\beta$.

To show this, we adopt a right-handed reference frame with the stellar spin axis along $\hat{\mathbf z}$. The observer’s inclination $i$ is the angle between the line of sight $\hat{\mathbf n}_{\rm obs}$ and $+\hat{\mathbf z}$,
\[
\hat{\mathbf n}_{\rm obs}=(\sin i,0,\cos i), \qquad 0\le i\le\pi,
\]
We denote $\beta$ the angle between the magnetic axis $\hat{\mathbf m}$ and $+\hat{\mathbf z}$. Thus, as the star rotates, the magnetic axis precesses around $\hat{\mathbf z}$:
\begin{equation}
\hat{\mathbf m}(\phi)
 =\big(\sin\beta\cos\phi,\ \sin\beta\sin\phi,\ \cos\beta\big),
\end{equation}
where $\phi$ is the rotational phase. Of the two rotational phases for which $\hat{\mathbf m}$ lies in the $\hat{\mathbf z}-\hat{\mathbf n}_{\rm obs}$ plane, we define $\phi=0$ as the one satisfying $\hat{\mathbf m}\cdot\hat{\mathbf n}_{\rm obs}>0$; the other corresponds to $\phi=\pi$. This choice of phase origin is a convenient convention and does not affect the physical conclusions.

All points on the emission ring where $B = B_{\rm loc}$ (Eq.~\ref{eq:Bloc}), are traced by the vector
\begin{equation}
\hat{\mathbf r}(\phi,\phi_m)\!
 = \!\sin\theta_m[\cos\phi_m\hat{\mathbf e}_1(\phi)\!
   +\! \sin\phi_m\hat{\mathbf e}_2(\phi)]\!
   +\! \cos\theta_m\hat{\mathbf m}(\phi).
\end{equation}
Here $\theta_m$ is the magnetic colatitude of the emission ring, and the unit vectors $\hat{\mathbf e}_1$ and $\hat{\mathbf e}_2$ span the plane perpendicular to $\hat{\mathbf m}(\phi)$.

\begin{figure}
\centering
\includegraphics[width=\linewidth,trim=100pt 0pt 50pt 0pt, clip]{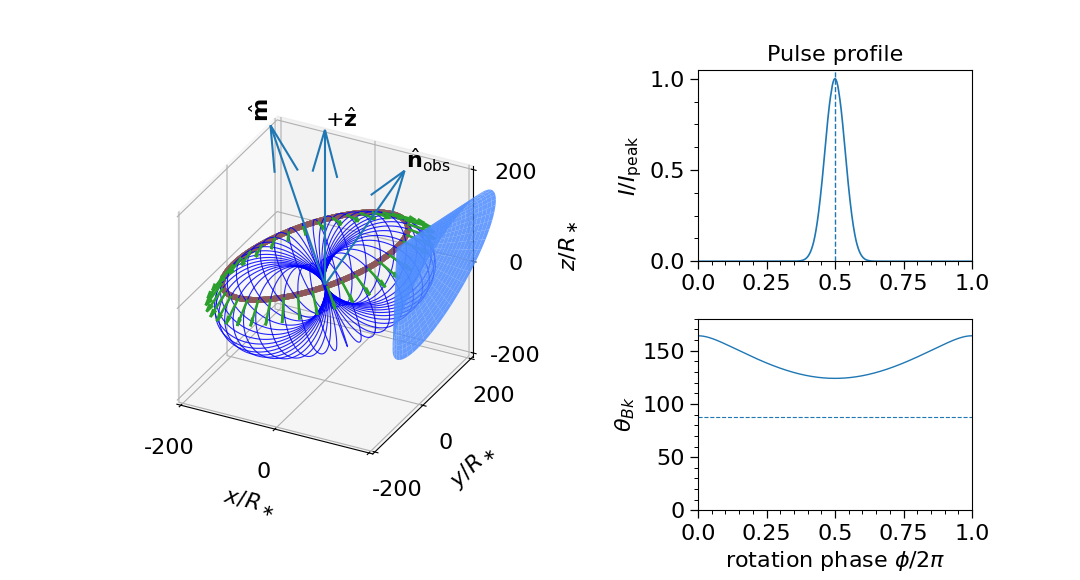}
\caption{Schematic of the ECME emission geometry in an oblique rotator. The spin axis $\hat{\mathbf z}$ defines the rotation frame, while the magnetic axis $\hat{\mathbf m}$ is tilted relative to it by an angle $\beta$. The observer’s line of sight $\hat{\mathbf n}_{\rm obs}$ makes an inclination $i$ to the spin axis. Magnetic field lines (depicted in blue) illustrate the dipolar topology. ECME arises where the local magnetic field equals the resonance value $B_{\rm loc}$. A representative hollow emission cone (shaded), centred on the local field direction $\hat{\mathbf b}_{\rm loc}$ and with opening angle $\theta_{\rm cone}$, is formed. For detected radiation, the wavevector $\mathbf{k}$ is aligned with $\hat{\mathbf n}_{\rm obs}$. As the star rotates, the hollow cone sweeps past the line of sight, producing one or two pulses per rotation depending on geometry. The right panels illustrate the resulting pulse profile and the phase variation of $\theta_{Bk}$.}
\label{fig:emission_cone}
\end{figure}

For a centred dipole, the local field direction is
\begin{equation}
\hat{\mathbf b}(\phi,\phi_m)
 = \frac{3(\hat{\mathbf m}\!\cdot\!\hat{\mathbf r})\hat{\mathbf r}-\hat{\mathbf m}}
        {\big\lVert 3(\hat{\mathbf m}\!\cdot\!\hat{\mathbf r})\hat{\mathbf r}
                   -\hat{\mathbf m}\big\rVert},
\end{equation}
and the instantaneous beaming angle is
\begin{equation}
\theta_{Bk}(\phi,\phi_m)
 = \arccos\!\big[\hat{\mathbf b}(\phi,\phi_m)\!\cdot\!\hat{\mathbf n}_{\rm obs}\big].
\end{equation}
The ECME is beamed into a hollow cone whose axis is aligned with the local magnetic-field direction $\hat{\mathbf b}_{\rm loc}$ and whose opening angle is $\theta_{\rm cone}$ (Fig.~\ref{fig:emission_cone}).

Emission is detectable when
\begin{equation}\label{eq:visibility}
|\theta_{Bk}(\phi,\phi_m)-\theta_{\rm cone}| \lesssim \sigma,
\end{equation}
as illustrated schematically in Fig.~\ref{fig:emission_cone}.

In the idealised case where the distribution of mirroring electrons is azimuthally uniform about the magnetic axis, and the ECME resonance condition is satisfied at \emph{all magnetic azimuths} characterised by $B=B_{\rm loc}$, ECME can operate at many azimuthal emission elements along the same field shell. Each azimuthal emission element produces a local hollow-cone beaming pattern in direction space, centred on its own local magnetic-field direction. At a given rotational phase, the observed radiation may arise from the intersection of the observer’s line of sight with one or several of the beaming surfaces of these local cones. Therefore, the pulse morphology of the radio emission (narrow, broad, or structurally rich pulses, as well as weaker leading and trailing components) depends on the viewing geometry $(i,\beta)$, the angular thickness of the cone wall $\sigma$, and the number of azimuthal emission elements whose beaming patterns intersect the line of sight, as well as the azimuthal extent over which this occurs.

The observed duty cycle $\Delta t/P$ is determined by how long the line of sight intersects the thin emission wall of the hollow cone as the star rotates. To first order,
\begin{equation}\label{eq:duty_sigma}
\frac{\Delta t}{P}
 \simeq \frac{\sigma}{\sin i\,\sin\zeta_{\rm loc}},
\end{equation}
where $\zeta_{\rm loc}$ is the angle between the spin axis and the local magnetic-field direction at the emitting field line. Very small duty cycles arise from thin cone walls ($\sigma\simeq0.3^\circ-0.5^\circ$) and geometries with $\sin i\,\sin\zeta_{\rm loc}\approx1$.

An antipodal magnetic pole may generate an interpulse separated by approximately half a rotation. Whether both hemispheres contribute to the observed radiation depends on the plasma distribution and its propagation toward the observer. Emission from one hemisphere can be strongly attenuated if it traverses a denser magnetised plasma column (see Section \ref{sec:polarisation}). Extending the geometric model to two poles simply requires adding the contribution from the antipodal emission region.

For a dipolar field and fixed harmonic, $\nu\propto B\propto r^{-3}$, so that lower-frequency emission originates at larger radii. This produces a small, systematic shift in the rotational phase at which the hollow cone intersects the line of sight. Thus, multi-band radio pulse phases probe both the surface field strength and the radial field gradient, since each observing band samples a different ECME resonance height set by $B(r)=B_{\rm loc}(\nu)$.

It is important to emphasise that the ECME visibility geometry differs fundamentally from the canonical pulsar picture. In pulsars, radio emission is typically modelled as a single phase-coherent beam associated with a global emission region, so detectability depends on whether this beam sweeps across the observer’s line of sight for a given $(i,\beta)$. Pulse formation then reflects the rotation of a filled or partially filled beam fixed in the magnetosphere. By contrast, ECME does not involve a single macroscopic beam. Coherent emission can occur simultaneously at many spatially separated resonant sites along a magnetic field shell where the local cyclotron condition is satisfied. Each site produces its own hollow-cone beaming pattern in \textit{direction space}, centred on the local magnetic-field direction. At a given rotational phase, observable radiation arises only if the line of sight intersects one or more of these beaming surfaces. Consequently, ECME pulse profiles are governed by how many resonant sites contribute at a given phase and over what azimuthal extent, rather than by the rotation of a fixed beam.

\section{Polarisation diversity in LPRTs under the ECME framework}\label{sec:polarisation}

We describe polarisation using the Stokes vector $\mathbf S=(I,Q,U,V)^\top$, where $I$ is total intensity, $Q$ and $U$ quantify linear polarisation, and $V$ quantifies circular polarisation with the sign of $V$ encoding handedness. The fractional linear and circular polarisations are $L/I=\sqrt{Q^{2}+U^{2}}/I$ and $|V|/I$, respectively.

LPRTs show pulses with both large $|V|/I$ and $L/I$, sometimes within the same burst. In the ECME picture, this arises from (i) intrinsically circular emission produced by maser growth at quasi-transverse propagation angles, and (ii) subsequent propagation through magnetised plasma above the ECME source that mixes Stokes parameters, as we illustrate below.

After generation, the radiation traverses a birefringent plasma. The transfer equations in Section \ref{sec:polarisedRT} show that the coefficients $\rho_R$ and $\rho_W$ govern Faraday rotation (mixing of $Q$ and $U$) and Faraday conversion (mixing of $U$ and $V$), respectively. If $s$ denotes the path length traversed by the emerging ray within the magnetised slab above the ECME source, the cumulative conversion depth 
\begin{equation}\label{eq:tau_C}
\tau_C=\int \rho_W\,ds
\end{equation}
quantifies how an initially circular state acquires linear power. We likewise define the rotation depth
\begin{equation}\label{eq:tau_R}
\tau_R=\int \rho_R\,ds,
\end{equation}
which measures the net Faraday rotation of the linear polarisation angle. For homogeneous conditions, $\tau_R\sim1$ corresponds to approximately a rotation of order half a radian.

\subsection{Polarised radiative transfer in Stokes form}\label{sec:polarisedRT}

We consider a plane-parallel slab, with the ray parameter $s$ measured along the line of sight, increasing in the direction of propagation from the ECME source ($s=0$) to the escape point ($s=s_{\rm esc}$). Let $\mathbf S_{\rm em}=(S_I,S_Q,S_U,S_V)^\top$ be the emissivity vector per unit path length. The transfer equation is
\begin{equation}\label{eq:stokes_transfer_general}
\frac{d\mathbf S}{ds}=\mathbf S_{\rm em}-\boldsymbol\eta\,\mathbf S,
\end{equation}
with propagation matrix
\begin{equation}\label{eq:propagation_matrix}
{\boldsymbol\eta}=
\begin{pmatrix}
\eta_I & \eta_Q & \eta_U & \eta_V\\
\eta_Q & \eta_I & -\rho_R & 0\\
\eta_U & \rho_R & \eta_I & -\rho_W\\
\eta_V & 0 & \rho_W & \eta_I
\end{pmatrix}.
\end{equation}
The diagonal coefficient $\eta_I$ describes isotropic attenuation of the total intensity $I$ along the propagation path (or amplification if $\eta_I<0$ in a maser). The coefficients $\eta_Q,\eta_U,\eta_V$ represent dichroic absorption, which couples the total intensity $I$ to the polarisation components $(Q,U,V)$. In particular, $\eta_V$ (circular dichroism) corresponds to differential absorption of right- and left-hand circular components and can generate circular polarisation from initially linear or unpolarised radiation, while $\eta_Q$ and $\eta_U$ (linear dichroism) represent differential absorption between orthogonal linear axes and generate linear polarisation with the electric vector position angle (EVPA) aligned to the favoured axis. The antisymmetric terms encode birefringent coupling among $Q,U,V$. The absence of a $\rho_U$ term follows from locally choosing the Stokes-$Q$ reference axis along the projection of $\mathbf B$ onto the sky. 

The antisymmetric part (the $\rho$ terms) in the propagation matrix (\ref{eq:propagation_matrix}) converts one type of polarisation into another without changing $I$. In the absence of attenuation (i.e., when all $\eta$ terms vanish) and when $\mathbf S_{\rm em}=0$, it preserves $P=(Q^2+U^2+V^2)^{1/2}$. The coefficient $\rho_R$ mixes $Q$ and $U$ (Faraday rotation of the EVPA), while the coefficient $\rho_W$ mixes $U$ and $V$ \citep[Faraday conversion, or the Voigt effect;][]{MartinWick1981_MagnetoOpticalEffectsI, MartinWick1982_MagnetoOpticalEffectsII}. In ECME, the mixing of $U$ and $V$ via $\rho_W$ during propagation causes the conversion that is chiefly responsible for transforming intrinsically circular polarisation into linear polarisation. From general principles, the transfer equation for polarised radiation must have this form and symmetry \citep{LandiInnocenti1973_RadiationTransfer_Stokes, MartinWick1979_SolutionsRadTransfer}.

ECME grows in a low-density region ($\omega_p/\Omega_e<0.1$), after which the amplified waves traverse a second region consisting of a magnetised plasma column. In this post-growth region, an initially circular ECME beam can emerge with substantial linear fractions, provided the cumulative conversion depth satisfies $\tau_C\sim0.1-1$.

We model the post-growth escape through this second region using a cold plasma slab of thickness $D$, in which the ray propagates quasi-transversely to the magnetic field, and we assume that dichroic attenuation is negligible. We set $\mathbf{S}_{\rm em}=0$, require $|\eta_{Q,U,V}|\,D\ll1$, and retain only the scalar attenuation $\eta_I$ and the birefringent terms. Under these conditions, the reduced Stokes system is
\begin{subequations}
\begin{align}
\frac{dI}{ds} &= - \eta_I I, \label{eq:reduced_system_a}\\
\frac{dQ}{ds} &= -\eta_I Q + \rho_R U,\label{eq:reduced_system_b}\\
\frac{dU}{ds} &= -\eta_I U - \rho_R Q + \rho_W V,\label{eq:reduced_system_c}\\
\frac{dV}{ds} &= -\eta_I V - \rho_W U.\label{eq:reduced_system_d}
\end{align}
\end{subequations}
In the cold-plasma, high-frequency limit ($\omega\gg\omega_p,\Omega_e$), the birefringent coefficients are \citep{Pacholczyk_RadioGalaxies}:
\begin{equation}\label{eq:rhos_cold}
\rho_R=\frac{\omega_p^2\,\Omega_e}{c\,\omega^2}\cos\theta_{Bk}, \qquad
\rho_W=\frac{\omega_p^2\,\Omega_e^{2}}{2c\,\omega^3}\sin^2\theta_{Bk},
\end{equation}
with $\cos\theta_{Bk}\equiv B_\parallel/B$ taken as signed, so the sign of $\rho_R$ is set by the line of sight field component $B_\parallel$. In this Stokes-transfer convention, $\rho_R$ corresponds to twice the rotation rate of the EVPA, $\chi$, i.e.\ ${\rm d}\chi/{\rm d}s=\rho_R/2$. These expressions apply only along the escape path outside the thin ECME growth layer and not inside the resonant maser zone ($\omega\simeq n\Omega_e$).

In quasi-transverse propagation ($\theta_{Bk}\approx90^\circ$), which is appropriate for LPRTs, the rotation coefficient $\rho_R$ becomes very small while the conversion coefficient $\rho_W$ remains finite. Faraday conversion therefore dominates over rotation, and neglecting both $\rho_R$ and attenuation gives the pure-conversion limit. With the $Q$ axis defined along the sky-projected $\mathbf B$, the transfer equations reduce to
\begin{equation}\label{eq:post_growth_pair_revised}
\frac{dQ}{ds}=0,\qquad 
\frac{dU}{ds}=\rho_W\,V,\qquad
\frac{dV}{ds}=-\rho_W\,U.
\end{equation}
These equations describe the mixing between $U$ and $V$ as the wave traverses the plasma slab. 

In the case of a small cumulative conversion depth
\begin{equation}
\tau_C \equiv \int_0^{s_{\rm esc}} \rho_W\,ds \ll 1,
\end{equation}
one obtains the following linearised behaviour at escape:
\[
U(s_{\rm esc})\simeq U_0+\tau_C V_0,\quad
V(s_{\rm esc})\simeq V_0-\tau_C U_0,
\]
so that an initially purely circular state ($U_0=0$, $V_0\simeq I$) acquires a small linear component $U/I\simeq\tau_C$. 

\begin{figure}
    \centering
    \includegraphics[width=0.98\linewidth,trim=320pt 0pt 0pt 0pt, clip]{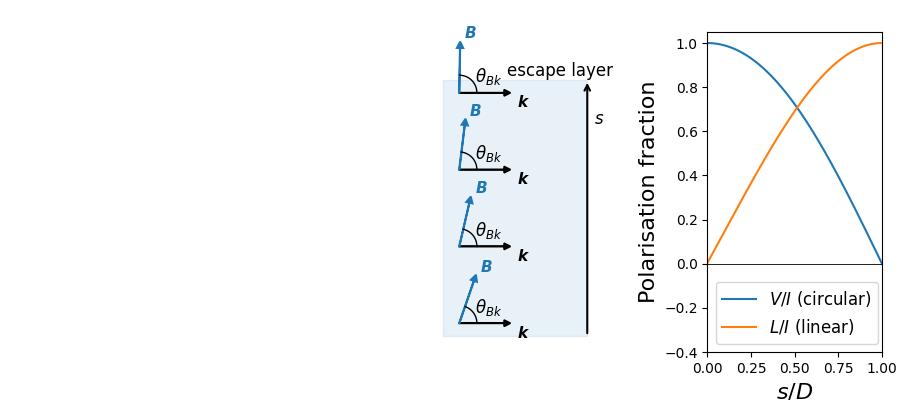}    \caption{Schematic illustration of propagation-induced circular to linear conversion as an ECME ray traverses an oblique magnetised slab of thickness $D$. The left panel shows the evolving field orientation relative to the propagation direction $\hat{\mathbf k}$, which denotes the local ray direction (tangent to the ray path) and is drawn with fixed orientation purely for visual clarity. The schematic is not drawn in physical coordinates: only the local angle $\theta_{Bk}(s)$ between $\hat{\mathbf k}$ and $\mathbf B$ is physically relevant. The right panel shows the corresponding change in $V/I$ and $L/I$ obtained by integrating the Stokes transfer equations, with $\theta_{Bk}(s)$ varying but remaining in the quasi-transverse regime required for ECME. For further details, see text.}
    \label{fig:schematic-propagation}
\end{figure}

In order to illustrate propagation-induced polarisation conversion in the ECME escape layer in a simplified manner, we have created a schematic diagram (see Fig.~\ref{fig:schematic-propagation}). In the left panel, a ray with wave vector $\hat{\mathbf k}$ (which denotes the local propagation direction, i.e., the tangent to the ray path) traverses a magnetised slab of thickness $D$. The local magnetic field $\mathbf B(s)$ is quasi-transverse to the ray (i.e., $\theta_{Bk}\simeq90^\circ$), and the angle $\theta_{Bk}(s)$ between $\mathbf B$ and $\hat{\mathbf k}$ increases monotonically along the path, representing the typical geometric evolution expected as the ray propagates through a curved dipolar field. For visual clarity, $\hat{\mathbf k}$ is drawn with a fixed orientation in the schematic, but the figure is not drawn in physical coordinates, and only the local angle $\theta_{Bk}(s)$ is physically meaningful. Thus, in this schematic picture, only the field orientation is allowed to vary.

In the right panel, we show the evolution of the fractional circular, $V/I$, and linear, $L/I$, polarisations obtained by integrating the Stokes transfer equations through the same slab for fixed $(n_e,B,\nu_{\rm obs})$, with $\theta_{Bk}(s)$ varying but remaining in the quasi-transverse regime required for ECME. The conversion coefficient scales as $\rho_W\propto n_e B^{2}\sin^{2}\theta_{Bk}$. In the schematic shown here, where $n_e$ and $B$ are held fixed, this reduces to $\rho_W\propto\sin^{2}\theta_{Bk}$, so a gradual increase in $\theta_{Bk}$ alone is sufficient to build up a conversion depth $\tau_C$ that transforms initially circular ECME into predominantly linear polarisation. In realistic magnetospheres, variations in both $n_e$ and $B$ also contribute to $\rho_W(s)$; however, isolating the variation of $\theta_{Bk}(s)$ alone provides the clearest demonstration of how propagation through a magnetised slab naturally produces circular to linear conversion.

On the other hand, in the regime with $\tau_C\sim1$, conversion is strong, and the emerging radiation can be predominantly linear even if the intrinsic maser emission was purely circular. We note that the conversion depth $\tau_C$ increases with electron density (via $\omega_p^2$), magnetic field strength (via $\Omega_e^2$), path length, and toward lower $\nu$ (as $\omega^{-3}$), reflecting the scaling of $\rho_W$ in the cold-plasma, high-frequency limit used here.

To illustrate the polarisation evolution, we now integrate the reduced Stokes system (Eqs.~\ref{eq:reduced_system_a}-\ref{eq:reduced_system_d}) for a uniform slab of thickness $D$, in which $\rho_W$ and $\rho_R$ are held constant along the ray (constant within each run, varied between runs), with $\mathbf{S}_{\rm em}=0$, $\eta_I D\ll1$, $\rho_W$ and $\rho_R$ taken from Eqs.~\eqref{eq:rhos_cold}, and $\theta_{Bk}=85^{\circ}$, so that $\rho_R$ is small but finite.

\begin{figure}
\centering
\includegraphics[width=\linewidth]{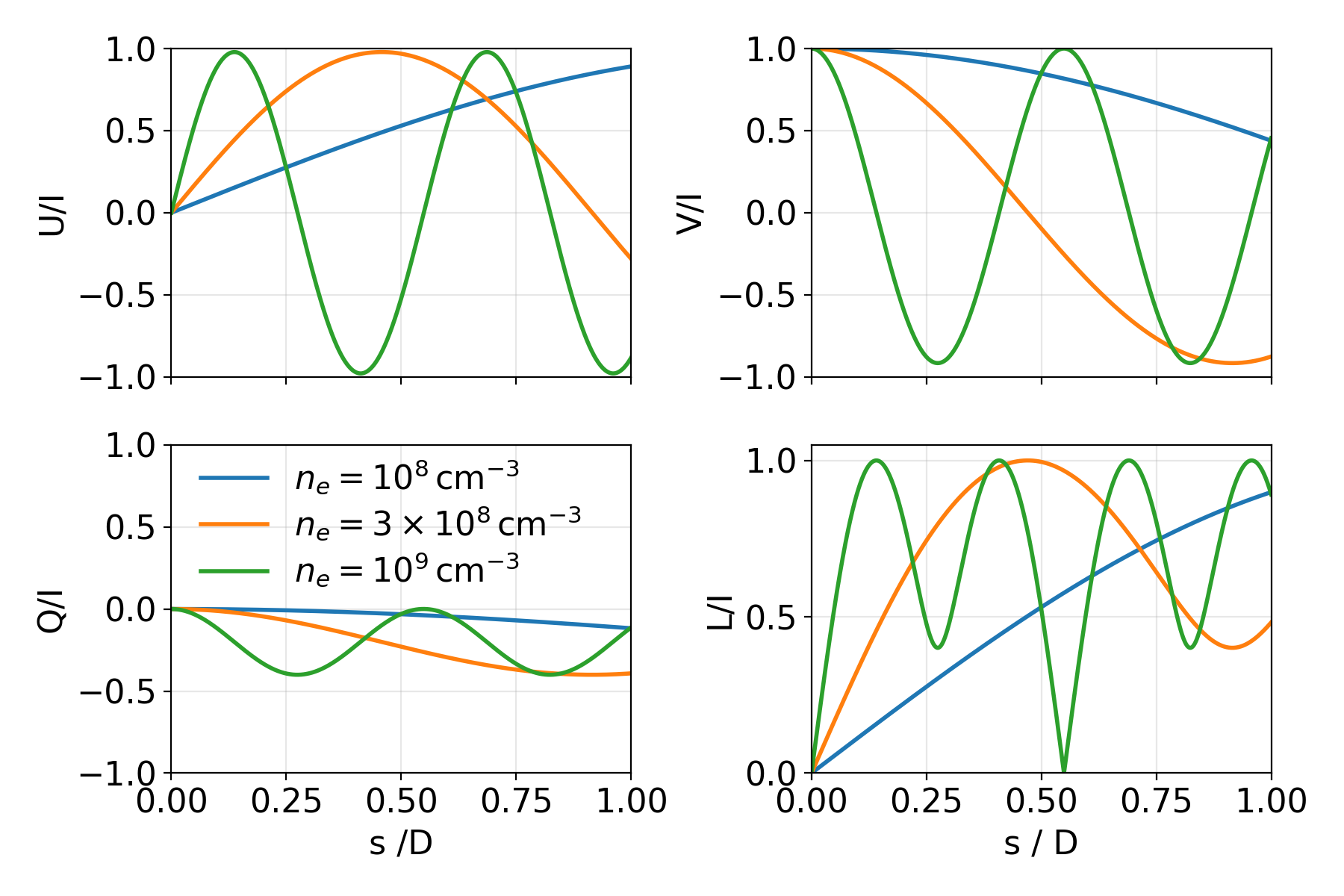}\\
\includegraphics[width=\linewidth]{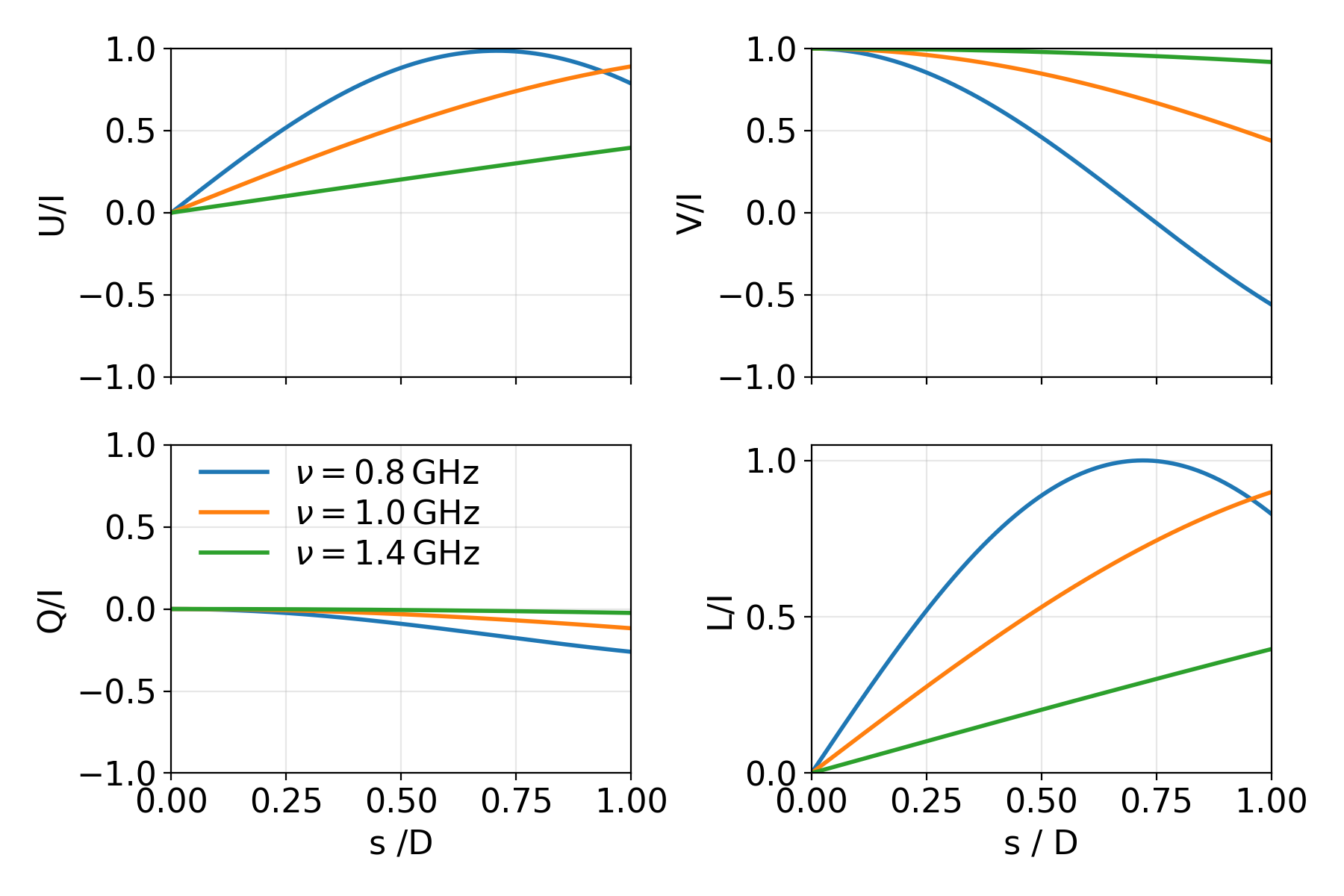}\\
\caption{Propagation induced evolution of the polarisation fractions in a cold, low absorption, quasi-transverse escape layer. Each panel shows $U/I$, $V/I$, $Q/I$, and $L/I=\sqrt{Q^{2}+U^{2}}/I$ as functions of $s/D$ for an initially circularly polarised state ($I_{0}=1$, $Q_{0}=U_{0}=0$, $V_{0}=1$).
\textbf{Top four panels:} Density sweep with $n_{e}=\{10^{8},\,3\times10^{8},\,10^9\}$\,cm$^{-3}$ at fixed $\nu_{\rm obs}=1$\,GHz and $\theta_{Bk}=85^{\circ}$. Both $\rho_{W}$ and $\rho_{R}$ scale linearly with $n_{e}$, so higher densities produce faster $U\leftrightarrow V$ oscillations and slightly larger $Q/I$ excursions. \textbf{Bottom four panels:} Frequency sweep with $\nu_{\rm obs}=\{0.8,\,1.0,\,1.4\}$\,GHz at fixed $n_{e}=10^{8}$\,cm$^{-3}$ and $\theta_{Bk}=85^{\circ}$. Increasing frequency weakens the mixing ($\rho_{W}\propto\nu^{-3}$, $\rho_{R}\propto\nu^{-2}$), yielding slower conversion, smaller $Q/I$ amplitudes, and reduced growth of $L/I$ at higher $\nu_{\rm obs}$.}
\label{fig:pol_sweeps}
\end{figure}

The initial conditions $(I_0,Q_0,U_0,V_0)=(1,0,0,1)$ represent 100 per cent circular ECME at the base of the propagation region, which correspond to $s=0$. For the runs shown in Fig.~\ref{fig:pol_sweeps} we vary $n_e$ and $\nu_{\rm obs}$ through their appearance in $\rho_W$ and $\rho_R$.

For LPRTs, the cumulative rotation depth $\tau_R$ (Eq.~\ref{eq:tau_R}) is typically much smaller than $\tau_C$ at GHz frequencies because $\rho_R\propto\omega^{-2}$ whereas $\rho_W\propto\omega^{-3}$, and because escape occurs close to $\theta_{Bk}\approx90^{\circ}$ where $\cos\theta_{Bk}$ (and hence $\rho_R$) is strongly suppressed (Eqs.~\ref{eq:rhos_cold}). Nevertheless, a small non-zero $\tau_R$ produces mild $Q/I$ excursions in the numerical solutions, as seen in Fig.~\ref{fig:pol_sweeps}, even though conversion still dominates the overall evolution of the polarisation fractions.

These results clearly illustrate how modest changes in the magnetised plasma drive the transition from predominantly circular to predominantly linear output.
To conclude, ECME does not prescribe a unique polarisation state. Instead, it can produce emission ranging from nearly purely circular to predominantly linear, set by the local density, the viewing geometry, and the characteristics of the magnetised plasma through which the escaping wave propagates.

\subsection{Application to LPRTs}\label{sec:lprt_application}

There are several classes of astrophysical sources that show highly polarised radio pulses that can be explained in terms of ECME. Some are binary systems, such as the so-called WD pulsars, whose prototype is AR~Scorpii \citep{Marsh2016_ARSco_Discovery, Stanway2018_ARSco_cyclotron}. In these systems, a MWD interacting with its companion exhibits coherent, strongly polarised emission. These WD pulsars are phenomenologically similar to MCVs, long recognised as ECME emitters \citep[for the first identification of ECME as a possible emission mechanism in MCVs, see][]{MelroseDulk1982_AMHerRadio_ECME}. These interacting magnetic binaries demonstrate that compact, magnetised stars can produce auroral radio beams \citep{Barrett2020AdSpR_RadioObs_MCVs_Maser}. However, these are well established binary systems that typically display obvious multiwavelength counterparts.

On the other hand, the LPRTs considered in this work are apparently isolated sources that possess periods of minutes to hours, have narrow radio duty cycles, show large and variable polarisation fractions, and, in most cases, lack a persistent optical/IR/X-ray counterpart. Phenomenologically, they share some properties with main-sequence radio pulse emitters \citep[MRPs,][]{DasChandra2021_MRP} in that the stable magnetosphere modulates a narrow beam powered by an electron cyclotron maser \citep[e.g., in CU Virginis][]{Trigilio2000_CUVir_RadioDiscovery, Ravi2010_RadioPulsusCUVir}. However, the energies of MRPs are several orders of magnitude lower than those of typical LPRTs, whose energetics seem to reflect the much deeper gravitational potential wells of compact stars.
\begin{table*}
\centering
\caption{
Observed properties of the LPRTs. Peak flux densities and observing frequencies refer to the brightest reported bursts. Distances are derived from Galactic electron-density models and may be uncertain by a factor of $\sim2$ \citep[e.g.][]{Yao2017_NewElectronDensityModelEstimationDistances}.}
\label{tab:lprt_objects}
\small
\begin{tabular}{@{}l c c c p{6.2cm}@{}}
\toprule
\textbf{Object} & \textbf{Period} & \textbf{Distance (kpc)} & \textbf{Multi-wavelength} & \textbf{Polarisation / radio behaviour} \\
\midrule

ASKAP\,J$1755-2527$\\
\begin{tabular}[l]{@{}l@{}}\citet{McSweeney2025_ASKAPJ175534-252749},\\ \citet{Dobie2024_ASKAPJ175534-252749}\end{tabular}
& 1.16 h
& 4.7
& Radio only
& Strong linear ($L/I\gtrsim60$ per cent) with substantial circular ($|V/I|\sim25-40$ per cent). PA behaviour varies between pulses. Highly intermittent. Duty cycle $\lesssim1$ per cent.
\\[3pt]

ASKAP\,J$1935+2148$\\
\citet{Caleb2024_LPRT_54min_Switching}
& 53.8 min
& $4.85$
& Radio only
& Three states: bright (linear $\sim90$ per cent, $10-50$\,s), weak (circular-dominated, $\sim370$\,ms), and quiescent. Duty cycle $0.3-1.5$ per cent.
\\[3pt]

GLEAM-X\,J$1627-5235$\\
\citet{Hurley-Walker2022_GLEAM-X1627}
& 18.18 min
& $1.3\pm0.5$
& Radio only
& Linear up to $\sim90$ per cent, no detected circular. Stable EVPA. Bursts $<0.5$\,s to $30-60$\,s. Duty cycle $0.9-2.7$ per cent.
\\[3pt]

GPM\,J$1839-10$\\
\citet{Hurley-WalkerRea2023_LPRT}, \\
\citet{MenMcSweeney2025_GPMJ1839-10}
& 21.97 min
& $5.7\pm2.9$
& Radio only
& Extreme variability. Linear from $\sim0$ to $\sim100$ per cent. Circular up to $\sim95$ per cent. $90^\circ$ PA jumps. When active, duty cycle $\sim5-10$ per cent, sometimes $>20$ per cent. Emission in only $\sim20$ per cent of rotations.
\\[3pt]

ASKAP/DART\,J$1832-0911$\\
\begin{tabular}[l]{@{}l@{}}\citet{Li2024_DARTJ1832-091_SNR},\\ \citet{WangRea2025_ASKAPDartJ1832-0911DetectionX-rays}\end{tabular}
& 44.2 min
& $\sim4.6$
& Radio and X-ray
& Linear up to $\sim100$ per cent with small phase-locked circular. Radio and X-ray pulses phase-locked. Wide-mode duty cycle $\sim4-11$ per cent. Rare ultra-narrow bursts also observed.
\\[3pt]

ASKAP\,J$1839-0756$\\
\citet{Lee2025_6.45hours_LPRT_Interpulses}
& 6.45 h
& $4.0\pm1.2$
& Radio only
& Main pulse: mixed linear (up to $\sim80$ per cent) and circular (up to $\sim40$ per cent) with sign change in $V$. Interpulse: linear $\sim90$ per cent. Pulse widths $320-710$\,s. Duty cycle $\sim1.4-3.1$ per cent.
\\[5pt]

\multicolumn{5}{@{}l}{\emph{Historic candidate:}} \\[3pt]

GCRT\,J$1745-3009$\\
\begin{tabular}[l]{@{}l@{}}\citet{Hyman2005_GCRTJ1745_GalCentre},\\ \citet{Roy2010A_GCRTJ1745_ECME}\end{tabular}
& 1.28 h
& $1-8.5$
& Radio only
& Circular up to $\sim100$ per cent. $\sim10$ min bursts every $\sim77$ min in discovery epoch. Highly sporadic. Duty cycle $10-13$ per cent. On/off emission modes.
\\
\bottomrule
\end{tabular}
\end{table*}

The observed diversity among LPRTs is fully consistent with the ECME framework. In what follows, we examine the sources listed in Table \ref{tab:lprt_objects} and interpret their observed properties within this framework.

\begin{itemize}
\item \textbf{ASKAP\,J$\mathbf{1755-2527}$} \citep{Dobie2024_ASKAPJ175534-252749, McSweeney2025_ASKAPJ175534-252749}: 
the original bright pulse exhibits a high degree of linear polarisation, reaching $L/I\gtrsim60$ per cent, together with a substantial circular component ($|V/I|\sim25-40$ per cent at peak), as shown by the full-Stokes dynamic spectra and polarisation profiles (Fig.~1 and the Stokes profiles in Figs.~2, 5 and 6 of \citealt{Dobie2024_ASKAPJ175534-252749}). Although the polarisation angle of this pulse shows a smooth rotation, subsequent pulses display flat or non-RVM-like PA behaviour (Fig.~4 of \citealt{McSweeney2025_ASKAPJ175534-252749}), where RVM refers to the rotating vector model in which the EVPA traces the projected orientation of a rotating magnetic dipole \citep{Radhakrishnan1969_RVM}. This indicates that the PA evolution is not a stable or reliable tracer of the underlying magnetic geometry, but instead varies between pulses. Such behaviour favours a significant role for propagation effects in shaping the observed polarisation, rather than a purely rotating-dipole origin. The coexistence of strong linear and circular polarisation is naturally explained if intrinsically circular ECME undergoes partial mode conversion in a magnetised plasma, with a moderate conversion depth ($\tau_C\sim0.1-1$) regulated by $\rho_W$. Pulse-to-pulse variability in PA and polarisation fractions can then plausibly arise from modest changes in plasma column density or magnetic-field orientation above the ECME source.
\item \textbf{ASKAP\,J$\mathbf{1935+2148}$} \citep{Caleb2024_LPRT_54min_Switching}: this source exhibits three distinct emission states. The bright-pulse state is characterised by highly linearly polarised emission, with pulse widths of $10-50$\,s and fractional linear polarisation approaching $\sim 90$ per cent \citep{Caleb2024_LPRT_54min_Switching}. The weak-pulse state consists of narrow ($\sim370$\,ms) pulses that are highly circularly polarised; for the MeerKAT detection the circular fraction exceeds 70 per cent while the linear fraction is $\sim40$ per cent \citep{Caleb2024_LPRT_54min_Switching}. The weak state is also about 26 times fainter than the bright state \citep{Caleb2024_LPRT_54min_Switching}. In addition, a quiescent (quenched) state is observed in which no radio pulses are detected, and the emission states evolve over month-long timescales, suggesting physical changes in the emission region \citep{Caleb2024_LPRT_54min_Switching}. Within an ECME framework, this phenomenology can be interpreted as arising from variations in magnetospheric plasma loading that shift the cumulative conversion depth $\tau_C$ and can intermittently suppress or restore observable maser emission.
\item \textbf{GLEAM-X\,J$\mathbf{1627-5235}$} \citep{Hurley-Walker2022_GLEAM-X1627}: the bursts show high linear polarisation (up to $\sim90$ per cent) and no detectable circular component within current sensitivity limits across a broad frequency range (Fig.~3 of \citealt{Hurley-Walker2022_GLEAM-X1627}). The EVPA is reported to be stable within and between bursts after RM correction, indicating emission from a locally ordered magnetic field region. Such nearly pure linear output can arise when the post-growth propagation occurs through plasma with conversion depths $\tau_C\gtrsim1$, so that intrinsic circular ECME is efficiently converted into linear. The bursts exhibit pronounced internal structure on short ($\lesssim0.5$\,s) timescales within overall burst durations of $30-60$\,s (Figs.~1 and 2 of \citealt{Hurley-Walker2022_GLEAM-X1627}). This behaviour can be attributed, in an ECME framework, to rotation sweeping the line of sight across different sections of the emission beam, with compact regions producing narrow bright features and broader regions yielding smoother emission. The overall behaviour is consistent with emission escaping through magnetospheric plasma in which strong Faraday conversion operates.
\item \textbf{GPM\,J$\mathbf{1839-10}$} \citep{Hurley-WalkerRea2023_LPRT, MenMcSweeney2025_GPMJ1839-10}: the source shows extreme pulse-to-pulse variability, with the fractional linear polarisation $L/I$ ranging from nearly zero to almost unity, while a strong circular component is often present and in some epochs reaches up to $\sim95$ per cent. 
The polarisation angle is generally flat across individual pulses but displays abrupt $90^\circ$ jumps on timescales of $0.2-4$\,s, and high-time-resolution polarimetry shows that these jumps coincide with rapid interconversion between linear and circular polarisation states \citep[Fig.~1 of][]{Hurley-WalkerRea2023_LPRT, MenMcSweeney2025_GPMJ1839-10}. This behaviour is explained if intrinsically circular ECME escapes through plasma with conversion depths $\tau_C\lesssim1$, so that small changes in plasma density or magnetic-field orientation shift the balance between $U$ and $V$. High-time-resolution observations also reveal fine temporal substructure and quasi-periodic features \citep[Extended Data Fig.~7 of][]{Hurley-WalkerRea2023_LPRT}.  MeerKAT follow-up observations detected emission in only 3 out of 15 consecutive rotations, demonstrating extreme rotation-to-rotation intermittency \citep{MenMcSweeney2025_GPMJ1839-10}.
\item \textbf{ASKAP/DART\,J$\mathbf{1832-0911}$} \citep{Li2024_DARTJ1832-091_SNR,WangRea2025_ASKAPDartJ1832-0911DetectionX-rays}: the radio emission shows both a broad, phase-locked mode with duty cycle of order $\sim4-11$ per cent and, on at least one occasion, an extremely narrow burst, together with strong polarisation variability, with pulse morphology and polarisation state changing between epochs. Phase-resolved Stokes profiles reveal a circularly polarised component that is locked to rotation phase in some epochs (see Stokes $V$ in Fig.~1 of \citealt{Li2024_DARTJ1832-091_SNR}), while a short (0.2\,s) FAST pulse detected during a radio-quiet interval is nearly 100 per cent linearly polarised with a flat position-angle curve (Fig.~3 of \citealt{Li2024_DARTJ1832-091_SNR}), corresponding to an extreme duty cycle of $0.007$ per cent. \citet{WangRea2025_ASKAPDartJ1832-0911DetectionX-rays} report the same radio and X-ray periodicity, with folded lightcurves showing a small and repeatable radio-X-ray phase offset, consistent with the radio and X-ray emitting regions maintaining a fixed geometric relationship within the rotating magnetosphere (Fig.~3 of \citealt{WangRea2025_ASKAPDartJ1832-0911DetectionX-rays}). The combination of episodic very high linear polarisation, a phase-locked circular component, and the coexistence of both broad and ultra-narrow duty-cycle emission is consistent with ECME escaping through plasma with $\tau_C\sim1$ and $\tau_R\ll1$. To produce the observed X-ray luminosity, the accretion rate (see Sec.~\ref{sec:longP_NS_ISM}) would need to be sufficiently high to power the X-ray hotspot and supply magnetospheric plasma, while remaining low enough that absorption does not suppress the escaping maser beam (i.e., $\eta_I D\ll1$).
\item \textbf{ASKAP\,J$\mathbf{1839-0756}$} \citep{Lee2025_6.45hours_LPRT_Interpulses}: the very long period ($P\simeq6.45$\,h), the presence of a bright main pulse and a weaker interpulse separated by $\simeq180^\circ$ in rotational phase, and the distinct EVPA-swing characteristics (Figs.~1 and 2 of \citealt{Lee2025_6.45hours_LPRT_Interpulses}) indicate an oblique or near-orthogonal rotator geometry in which emission is observed from both magnetic hemispheres. The main pulse exhibits mixed linear and circular polarisation with a sign reversal in Stokes $V$ across the pulse window (Fig.~2 of \citealt{Lee2025_6.45hours_LPRT_Interpulses}). In an ECME framework, the intrinsic emission is circularly polarised, with the handedness set by the projection of the local magnetic field onto the propagation direction; as the line of sight sweeps across different azimuths of the hollow emission cone, this projection can change sign across the pulse window. For partial mode conversion ($\tau_C\sim1$), a residual circular component survives, imprinting this geometrical reversal in Stokes $V$. In contrast, the interpulse is almost purely linearly polarised with negligible circular component (Fig.~2 of \citealt{Lee2025_6.45hours_LPRT_Interpulses}), consistent with a viewing geometry or local plasma conditions in which conversion is strong ($\tau_C\gg1$). The main-pulse widths of $320-710$\,s correspond to a duty cycle of $\simeq1.4-3.1$ per cent, comparable to other LPRTs.
\item \textbf{GCRT\,J$\mathbf{1745-3009}$} \citep[historic candidate,][]{Hyman2005_GCRTJ1745_GalCentre, Roy2010A_GCRTJ1745_ECME}: the source was originally discovered as a series of $\sim1$\,Jy, $\sim10$\,min bursts recurring on a $\sim77$\,min timescale at $\nu\simeq0.33$\,GHz (Figs.~1 and 2 of \citealt{Hyman2005_GCRTJ1745_GalCentre}). A reanalysis of a 2003 GMRT detection shows strongly time-variable circular polarisation, with Stokes $V$ reversing sign and the fractional circular polarisation approaching $\sim100$ per cent (Fig.~2 of \citealt{Roy2010A_GCRTJ1745_ECME}). In our propagation framework, the near-unity circular fraction corresponds to weak conversion along the escape path ($\tau_C\ll1$), while the sign reversal indicates a change in the projection of the local magnetic field onto the propagation direction during the burst. On the basis of the extreme circular polarisation and high inferred brightness temperature, \citet{Roy2010A_GCRTJ1745_ECME} argue that the emission is coherent and identify ECME or plasma emission as the most plausible origin scenarios.
\end{itemize}

Taken together, these sources demonstrate that the wide phenomenological diversity of LPRTs is naturally accommodated within the ECME framework once propagation effects are included.

\section{Interaction with the environment}\label{sec:longP_NS_ISM}

LPRTs require an explanation for (i) how compact stars can attain spin periods of minutes to hours, and (ii) how they maintain the non-thermal electron populations needed for coherent emission.

Both MWDs and neutron stars are, in principle, capable of producing ECME. In the case of MWDs, spin periods of minutes to hours are largely inherited through their evolutionary history. For neutron stars, however, an additional mechanism is required to account for such long rotational periods, since vacuum dipole braking alone cannot spin pulsars down to periods of minutes to hours within Galactic timescales.

In Section~\ref{sec:origin_longP}, we examine how magnetospheric torques arising from captured ISM material can drive neutron stars from ordinary pulsar periods to $P\sim10^{2}-10^{4}$\,s. The role of the same ISM inflow in supplying the particles and energy required to sustain ECME is discussed in Section~\ref{sec:injection} for both neutron stars and MWDs.

\subsection{Origin of long-period neutron stars}\label{sec:origin_longP}

Neutron stars are typically born with initial periods $P_{0}\sim0.03-0.3$\,s and $B_{\ast}\sim10^{11}-10^{13}$\,G, so attaining periods of minutes to hours requires additional spin-down.

Vacuum dipole braking alone is insufficient. Under the standard dipole assumptions of braking index $n=3$, constant $B_{\ast}$, negligible initial period ($P_{0}\ll P$), $I=10^{45}$\,g\,cm$^{2}$, $R_{\ast}=10^6$\,cm, and $B_{\ast}=10^{12}$\,G, the dipole spin-down timescale becomes
\begin{equation}
\tau_{\rm dip}\approx1.7\times10^{13}
\left(\frac{P}{10^{3}\,{\rm s}}\right)^{2}\,{\rm yr},
\end{equation}
so reaching $P\sim10^{3}$\,s would exceed a Hubble time.

A faster channel is early supernova fallback \citep[e.g.][]{Alpar2001_PulsarSpinDown_FallbackDisk}. Accretion from a disc or torus can spin the neutron star down from seconds to $10^{3}-10^{4}$\,s within $10^{3}-10^{5}$\,yr, while the external dipole is still relatively strong and simultaneously bury it to $B_\ast\sim10^{8}-10^{10}$\,G. We will see in Section~\ref{sec:injection} that such fields favour ECME because lower surface fields place the GHz resonance at larger radii, where they more readily satisfy $\omega_{p}/\Omega_{e}\lesssim0.3$ in the maser-emission region.

A slower but still viable route is long-term interaction with the ISM \citep[e.g.][]{ProkhorovPopovKhoperskov2002Period_OldNS}. A neutron star of mass $M$ and velocity $v$ moving through gas of electron density $n_{e}$ captures material at the rate
\begin{equation}\label{eq:BHLMdot_short}
\dot M \simeq 4\pi G^{2} M^{2} m_{p}\, n_{e}\, v^{-3},
\end{equation}
where $m_p$ is the proton mass and $n_e$ denotes the effective electron density of the plasma coupled to the magnetosphere. For the purposes of magnetospheric interaction, we parametrise the captured mass density as $\rho \simeq m_p n_e$.

For $M=1.4\,M_{\odot}$ the corresponding capture rate is
\[
\dot M \simeq 7.3\times10^{8}
\left(\frac{n_{e}}{1\,{\rm cm^{-3}}}\right)
\left(\frac{v}{100\,{\rm km\,s^{-1}}}\right)^{-3}\,{\rm g\,s^{-1}},
\]
so for $n_{e}=1-10$\,cm$^{-3}$ we obtain $\dot M\simeq(0.7-7.3)\times10^9$\,g\,s$^{-1}$.

If the Alfv\'en radius exceeds the corotation radius,
\begin{equation}
R_A > R_{\rm co}, \qquad
R_{\rm co}=\left(\frac{GM}{\Omega^2}\right)^{1/3},
\end{equation}
the inflowing gas is centrifugally expelled and the rotating magnetosphere applies a propeller-like torque that extracts angular momentum from the star,
\begin{equation}\label{eq:prop_torque_short}
I\dot\Omega \simeq -\dot M\,\Omega\,R_{A}^{2}.
\end{equation}
For Bondi-Hoyle capture from the ISM, the Alfv\'en radius is
\begin{equation}\label{eq:RA_scaling_ne}
\begin{aligned}
R_{A}\simeq{}&3.2\times10^{10}
\left(\frac{B_{\ast}}{10^{12}\,{\rm G}}\right)^{4/7}
\left(\frac{R_{\ast}}{10^6\,{\rm cm}}\right)^{12/7}
\\[2pt]
&\times
\left(\frac{M}{1.4\,M_{\odot}}\right)^{-5/7}
\left(\frac{n_{e}}{1\,{\rm cm^{-3}}}\right)^{-2/7}
\left(\frac{v}{100\,{\rm km\,s^{-1}}}\right)^{6/7}
\,{\rm cm}.
\end{aligned}
\end{equation}
The spin evolution follows
\begin{equation}\label{eq:spin_evolution}
\Omega(t)=\Omega_{0}\,{\rm e}^{-t/\tau_{\rm prop}}
\end{equation}
where
\begin{equation}\label{eq:tau_prop}
\tau_{\rm prop}=\frac{I}{\dot M R_{A}^{2}}
\end{equation}
Using Eqs.~\eqref{eq:BHLMdot_short} and \eqref{eq:RA_scaling_ne}, the characteristic timescale becomes
\begin{equation}\label{eq:tau_prop_scaling_numeric}
\begin{aligned}
\tau_{\rm prop}\approx{}&4.2\times10^7
\left(\frac{I}{10^{45}\,{\rm g\,cm^{2}}}\right)
\left(\frac{B_{\ast}}{10^{12}\,{\rm G}}\right)^{-8/7}
\left(\frac{R_{\ast}}{10^6\,{\rm cm}}\right)^{-24/7}
\\[2pt]
&\times
\left(\frac{M}{1.4\,M_{\odot}}\right)^{-4/7}
\left(\frac{n_{e}}{1\,{\rm cm^{-3}}}\right)^{-3/7}
\left(\frac{v}{100\,{\rm km\,s^{-1}}}\right)^{9/7}
\,{\rm yr}.
\end{aligned}
\end{equation}
For the fiducial parameters $B_{\ast}=10^{12}$\,G, $M=1.4\,M_{\odot}$, $R_{\ast}=10^6$\,cm, $I=10^{45}$\,g\,cm$^{2}$, and $n_{e}=1-10$\,cm$^{-3}$, this gives
\[
\tau_{\rm prop}\approx(1.6-4.2)\times10^7\,{\rm yr}.
\]
For the same parameters, Eq.~\eqref{eq:RA_scaling_ne} gives $R_{A}\simeq(1.7-3.2)\times10^{10}$\,cm.

Integrating Eq.~\eqref{eq:spin_evolution} yields the time to evolve from $P_{0}=0.5$\,s to $P$:
\begin{equation}
t=\tau_{\rm prop}\,\ln\left(\frac{P}{P_{0}}\right),
\end{equation}
so achieving $P=10^{3}-10^{4}$\,s requires $t\sim(1-4)\times10^{8}$\,yr, well within the age of the Galaxy.

\subsection{Injection of non-thermal electrons and the ISM power supply}\label{sec:injection}

Coherent radio bursts in LPRTs require a persistent supply of mildly relativistic electrons; a natural and sustainable source of such particles is weak accretion from the surrounding ISM. While the discussion here focuses on neutron stars, we stress that the same particle-supply arguments apply to MWDs. 

Accretion from the ISM onto old, isolated neutron stars was first proposed by \citet{Ostriker1970_OldNeutronStars_ISM} and later revisited by \citet{TrevesColpi1991_AccretionISM} and \citet{BlaesMadau1993_accretion_ISM} as a possible source of faint soft X-ray emission detectable with the \textit{ROSAT} All-Sky Survey. However, no secure population of such objects emerged \citep{Colpi1998_OldNeutronStars_ISM}. Subsequent work demonstrated that the classical Bondi-Hoyle estimate systematically overpredicts the luminosity because it neglects magnetic and rotational effects that strongly inhibit inflow \citep[e.g.][]{Perna2003_missingNS_ISM}. MHD simulations confirm that the magnetosphere can substantially reduce the accretion rate relative to the Bondi-Hoyle value, either through magnetic pressure or propeller action \citep{Romanova2003_MHDpropeller, ToropinaRomanova2012_MHDAccretionISM_NS}. 

Nonetheless, the quasi-simultaneous radio and X-ray pulses detected from ASKAP/DART\,J$1832-0911$  suggest that an ISM-driven accretion channel may operate in some isolated neutron stars. The same systems may therefore manifest as LPRTs, with coherent ECME powered by the weak inflow that also produces X-rays. This provides an observationally supported and physically self-consistent energy pathway. In this regime, the captured material is funnelled by the stellar magnetic field onto small polar caps, where it is thermalised and emits X-rays. At the same time, a fraction of the inflowing electrons is accelerated along converging field lines, maintaining the pitch-angle anisotropy required for maser amplification. Because the X-rays originate near the magnetic footprints of closed field lines, while the radio emission arises higher up where $B=B_{\rm loc}$, the two bands are expected to be nearly phase-aligned, so that the same ISM inflow powers both the soft X-ray component and the ECME bursts.

For a compact object of mass $M$ and radius $R_\ast$ moving at velocity $v$ through interstellar gas of electron density $n_e$, the Bondi-Hoyle upper bound on the gravitational accretion power is
\begin{eqnarray}\label{eq:Lacc}
L_{\rm acc} &\simeq& \frac{GM\dot M}{R_\ast} 
\simeq 1.3\times10^{29}
\left(\frac{M}{1.4\,M_\odot}\right)^{3}
\left(\frac{R_\ast}{10^6\,\mathrm{cm}}\right)^{-1}\nonumber\\
& & \qquad \times 
\left(\frac{n_e}{1\,\mathrm{cm^{-3}}}\right)
\left(\frac{v}{100\,\mathrm{km\,s^{-1}}}\right)^{-3}
\mathrm{erg\,s^{-1}}.
\end{eqnarray}
For typical interstellar densities $n_e\sim1-10$\,cm$^{-3}$ and velocities $v\sim30-100$\,km\,s$^{-1}$ (corresponding to the low-velocity tail of the neutron-star population), the Bondi-Hoyle estimate gives $L_{\rm acc}\sim10^{29}-10^{31}$\,erg\,s$^{-1}$. Magnetic pressure and rotational inhibition reduce the actual inflow by factors of $10^{-2}-10^{-3}$ \citep{BlaesMadau1993_accretion_ISM, ToropinaRomanova2012_MHDAccretionISM_NS}, leaving an effective luminosity of up to $\sim 10^{29}$\,erg\,s$^{-1}$. This is sufficient to power ECME detectable to kiloparsec distances, because the emission is coherent and strongly beamed.  Although the associated X-ray luminosities would typically fall below the detection thresholds of all-sky soft X-ray surveys for sources located beyond a few hundred parsecs, particularly once interstellar absorption is included, neutron stars, and thus also LPRTs, could occasionally reveal themselves as X-ray emitters at times when the ambient gas density and relative velocity (see sections \ref{sec:velocity_ceiling} and \ref{sec:discussion}) can support short-lived episodes of enhanced accretion.

\subsection{Power budget and a velocity ceiling}\label{sec:velocity_ceiling}

Let $f_b$ be the beaming fraction, $\eta_r$ the efficiency for converting the total accretion power into coherent radio emission, and $L$ the isotropic-equivalent radio luminosity inferred from the observed bursts. The true emitted radio power is $f_bL$, thus requiring
\begin{equation}
L_{\rm acc}\ \ge\ \frac{f_b\,L}{\eta_r},
\label{eq:power_condition}
\end{equation}
which, combined with Eq.~\eqref{eq:Lacc}, yields an explicit upper limit on the stellar velocity:
\begin{equation}\label{eq:v_limit}
\begin{aligned}
v &\le 100
\left[\frac{1.3\times10^{29}\,\eta_r\,n_e}
{f_b\,L}\right]^{1/3}\,{\rm km\,s^{-1}}\simeq 35
\left(\frac{L}{10^{31}\,{\rm erg\,s^{-1}}}\right)^{-1/3}\!\!\! \\
&
\times \left(\frac{f_b}{0.03}\right)^{-1/3}\!\!\!
\left(\frac{\eta_r}{0.1}\right)^{1/3}\!\!
\left(\frac{n_e}{1\,{\rm cm^{-3}}}\right)^{1/3}\!
\,{\rm km\,s^{-1}}.
\end{aligned}
\end{equation}
For representative values $L=10^{31}$\,erg\,s$^{-1}$ (the isotropic-equivalent peak radio luminosity), $f_b=0.03$ (appropriate for a narrow ECME hollow cone with a thin wall), and $\eta_r=0.1$ (a plausible efficiency for coherent maser emission), Eq.~\eqref{eq:v_limit} yields $v\lesssim35\,n_e^{1/3}$\,km\,s$^{-1}$, i.e., $\sim35$\,km\,s$^{-1}$ for $n_e=1$\,cm$^{-3}$ and $\sim75$\,km\,s$^{-1}$ for $n_e=10$\,cm$^{-3}$.

Hence, only slowly moving neutron stars embedded in moderately dense interstellar regions can satisfy the power budget.

\subsection{Accretion versus propeller}\label{sec:accretion_propeller}

Whether accretion proceeds to the stellar surface or is inhibited by the propeller effect depends on the relative positions of the Alfvén radius $R_A$ (Eq.~\ref{eq:RA_scaling_ne}) and corotation radius $R_{\rm co}=\left({GM}/{\Omega^{2}}\right)^{1/3}$. Accretion occurs when $R_A<R_{\rm co}$.
Because $\dot M\propto n_{e}v^{-3}$, this condition defines an upper limit to $B_{\ast}$ for given $(P,n_{e},v)$. For $B_{\ast}\sim10^{8}-10^{10}$\,G and $P\sim10^{3}-10^{4}$\,s, one finds $R_{A}<R_{\rm co}$ over a broad region of parameter space ($n_{e}=1-10$\,cm$^{-3}$, $v\lesssim80$\,km\,s$^{-1}$), allowing matter to reach the surface and release upper-bound accretion luminosities of $L_{\rm acc}\sim10^{29}-10^{31}$\,erg\,s$^{-1}$.

For higher magnetic fields ($B_{\ast}\gtrsim10^{12}$\,G) at comparable spin periods, $R_{A}>R_{\rm co}$ and accretion is centrifugally inhibited (propeller regime). In this case, the maximum available gravitational power is limited to the work done near $R_{A}$:
\begin{equation}\label{eq:Lmax_propeller}
\begin{aligned}
L_{\rm max} \simeq \frac{GM\dot M}{R_{A}}
\simeq  3\times10^{28} &
\left(\frac{B_{\ast}}{10^{12}\,{\rm G}}\right)^{-4/7}
\left(\frac{n_{e}}{1\,{\rm cm^{-3}}}\right)^{9/7}\\
&\times \left(\frac{v}{100\,{\rm km\,s^{-1}}}\right)^{-27/7}
\,{\rm erg\,s^{-1}}.
\end{aligned}
\end{equation}
Allowing for a leakage fraction $f_{\rm leak}\lesssim10^{-3}$, representing the small portion of inflowing material that penetrates the centrifugal barrier through magnetic reconnection or Kelvin-Helmholtz instabilities at the magnetospheric boundary \citep[e.g.][]{Romanova2003_MHDpropeller}, gives an effective luminosity $L_{\rm eff}\simeq f_{\rm leak}L_{\rm max}\sim10^{26}-10^{27}$\,erg\,s$^{-1}$, which is insufficient, on energetic grounds, to sustain detectable ECME or X-ray emission beyond a few hundred parsecs if  $B_{\ast}\gtrsim 10^{12}$\,G. 

Even before the strict propeller condition $R_A>R_{\rm co}$ is satisfied, the available accretion power decreases rapidly with increasing surface magnetic field strength. From Eq.~\eqref{eq:Lmax_propeller}, $L_{\rm max}\propto B_\ast^{-4/7}$, so that increasing $B_\ast$ from $10^{10}$ to $10^{11}-10^{12}$\,G reduces the maximum gravitational power by more than an order of magnitude for fixed $(n_e,v)$. When combined with realistic leakage fractions ($f_{\rm leak}\ll1$), this steep scaling implies that neutron stars with $B_\ast\gtrsim10^{10}$\,G are increasingly unable to supply the $\sim10^{28}-10^{29}$\,erg\,s$^{-1}$ required to sustain ECME detectable at kiloparsec distances under typical ISM conditions. Detectability, therefore, strongly favours slowly rotating neutron stars with comparatively low surface magnetic fields, $B_\ast\sim10^8-10^{10}$\,G.

\subsection{ECME viability and the neutron star versus MWD central engine}

Taken together, Sections~\ref{sec:injection}, \ref{sec:velocity_ceiling}, and \ref{sec:accretion_propeller} show that the most plausible ECME-powered LPRTs are  long-period, relatively low-field neutron stars moving slowly through moderately 
dense interstellar gas. In this framework, ISM accretion can supply both the non-thermal electron population required for ECME and, under favourable conditions, a small thermal hotspot at the magnetic footprints. While such X-ray emission would generally be faint and often undetectable, the same  field-line geometry naturally allows radio and X-ray pulses to appear nearly phase-aligned when both are observable, without invoking a binary companion. The detection of phase-aligned radio and X-ray pulses from ASKAP/DART\,J$1832-0911$ provides a concrete example that is consistent with this picture.

The alternative possibility is that some LPRTs might instead be powered by isolated MWDs.  These objects possess strong magnetic fields  \citep[$B_{\ast}\simeq10^{4}-10^9$\,G][]{FerrarioWickKawka2020_MWDs} and have extended magnetospheres capable of sustaining ECME. However, they differ sharply from neutron stars in accretion power, curvature radii, and achievable luminosities. We, therefore, examine whether a MWD can satisfy the same geometric and energetic requirements.

The much larger magnetospheres of MWDs yield lower plasma densities at comparable inflow rates, thus making the maser condition $\omega_{p}/\Omega_{e}\lesssim0.3$ easier to satisfy. The difficulty lies in the available power: even under optimistic ISM accretion, a MWD with $B_{\ast}\sim10^{8}$\,G, $M\simeq0.8\,M_{\odot}$, and $R_{\ast}\simeq10^9$\,cm can only supply $L_{\rm acc}\sim10^{26}-10^{27}$\,erg\,s$^{-1}$, below the isotropic-equivalent radio luminosities inferred for LPRTs (see Eq.~\ref{eq:Lacc}). Thus, while ECME may occur, the emission from MWDs would be too weak to be detected beyond a few hundred parsecs.

Additionally, in a dipolar magnetic field, the local radius of curvature quantifies how sharply a field line bends in the meridional plane. It is given by \citep{Gangadhara2004_Radius_curvature}
\begin{equation}
R_{\rm c}=\frac{r(1+3\cos^{2}\theta)^{3/2}}{3\sin\theta(1+\cos^{2}\theta)},
\end{equation}
which reduces to $R_{\rm c}\simeq r/3$ at the magnetic equator. A smaller $R_{\rm c}$, therefore, corresponds to a more tightly curved field line and a faster change in the local magnetic-field direction along that line. At fixed resonance field $B_{\rm loc}$, one has $R_{\rm c}\propto R_{\ast}B_{\ast}^{1/3}$,  so any neutron star possesses a curvature radius orders of magnitude smaller than that of a MWD. 

To give some illustrative examples, for ECME at $\nu_{\rm obs}\simeq1$\,GHz ($B_{\rm loc}\simeq357$\,G) in a MWD with $B_{\ast}=10^{8}$\,G, the resonance condition places the emitting layer at $r\simeq6.5\times10^{10}$\,cm, giving $R_{\rm c}\sim2.2\times10^{10}$\,cm. In comparison, a neutron star with $B_{\ast}=10^{12}$\,G would host the same resonance at $r\simeq1.4\times10^9$\,cm, yielding $R_{\rm c}\simeq5\times10^{8}$\,cm, about fifty times smaller. 

For a neutron star with a weaker field of $B_\ast = 10^8$ G, the same scaling gives $R_c\simeq2.2\times10^7$ cm, nearly $10^{3}$ times smaller than for MWDs, simply because of the much smaller neutron star radius, reinforcing the expectation of strong geometric rotational modulation and making such low-field neutron stars even more compelling candidates for LPRTs (see below).

The much larger curvature radius in MWDs means that, for small $\beta$, the field orientation changes only slowly across the emission region. As a result, the beaming direction varies little during a full rotation, allowing several field-line bundles, in the idealised case, to satisfy the maser resonance condition simultaneously if the angle between the field and line of sight is favourable. Emission from these extended auroral arcs can therefore overlap in rotational phase, producing a quasi-steady radio signal. For MWDs with large magnetic obliquities ($\beta\gtrsim60^{\circ}$), however, the field direction at the ECME layer changes significantly as the star rotates, leading to stronger amplitude modulation even if the emission is intrinsically continuous. 

Finally, dedicated radio searches have found no evidence for coherent emission from isolated MWDs. A wide VLASS cross-match shows that, within current survey sensitivities, strong ($\gtrsim1-3$\,mJy at 3\,GHz) radio emission, whether continuous or pulsed, is virtually absent outside interacting binaries hosting a MWD, while the most sensitive targeted observations of isolated MWDs yield only non-detections with stringent upper limits \citep{Pelisoli2024_VLASS_WD_Survey, Zhang2025_IsolatedWD_RadioLimits}.

In close binaries containing a MWD, magnetic interaction with the companion can create multiple maser sites with differing field orientations and time-variable beaming due to orbital motion and asynchronous rotation. These effects enlarge the effective beaming solid angle and make ECME far more detectable, consistent with the comparatively higher incidence of coherent radio emission among MWDs in interacting systems.

\section{Discussion}\label{sec:discussion}

ECME has not been previously developed as a general framework for the class of isolated LPRTs. On the other hand, old and isolated neutron stars have long been expected to emit soft X-rays through accretion from the ISM \citep{Ostriker1970_OldNeutronStars_ISM}. 

The discovery of LPRTs may have therefore revealed a previously unrecognised long-$P$, low-$B_{\ast}$ regime in which ECME can operate efficiently, allowing isolated neutron stars to be identified and studied as radio sources and, occasionally, as X-ray emitters when the inflow from the environment provides sufficient power. In this context, ASKAP/DART\,J$1832-0911$ may reside in a comparatively dense region of the inner Galactic plane, possibly at the outskirts of a supernova remnant or superbubble, where the ambient gas density and relative velocity can support short-lived episodes of enhanced accretion. The coexistence of ECME bursts and moderate X-ray luminosity is therefore not contradictory: ECME is powered by mildly relativistic electrons in the outer magnetosphere of the neutron stars where $B_{\rm loc}\sim10^{2}-10^{3}$\,G, whereas the X-rays arise from intermittent accretion onto the stellar surface. Thus, these systems provide a new diagnostic of their surroundings, since the radio luminosity, duty cycle, and occasional X-ray emission depend sensitively on the local ambient density, ionisation state, and relative velocity. The Galactic latitude distribution of LPRTs reported by \citet{Dobie2024_ASKAPJ175534-252749} (called Ultralong Period sources by these authors) is fully consistent with the expectations of our ECME model requiring neutron stars to move slowly through the ISM and to possess magnetic fields $B_\ast\lesssim10^{10}$\,G. In this picture, LPRTs need not represent a distinct class of neutron stars, but rather a subset drawn from the low-velocity and low-field tail of the broader neutron star population. Classical radio pulsars typically occupy higher-field and higher-velocity regions of these distributions, but magnetic field decay can naturally allow older neutron stars to evolve into the parameter space favourable for ECME. A clear prediction of this framework is that isolated LPRTs should not be found at large heights above the Galactic plane, where the ambient ISM is too rarefied to supply the magnetospheric plasma required for ECME. Consistent with this expectation, the few LPRTs located at $|z|\sim300-400$\,pc, such as ILT\,J1101$+$5521 and GLEAM-X\,J0704$-$37, are known binary systems, for which the plasma required for coherent emission can be supplied by a companion rather than the diffuse ISM.

To place the proposed propagation scenario on a quantitative basis, it is useful to clarify the physical nature of the magnetised ``slab'' responsible for Faraday conversion. For slowly moving, low-field neutron stars, Bondi-Hoyle capture from an ambient ISM with density $\sim0.1-10$\,cm$^{-3}$ can supply a tenuous but magnetised plasma column with electron densities $n_e\sim10^{2}-10^{6}$\,cm$^{-3}$ in the outer magnetosphere. For the low-field neutron stars favoured in our framework ($B_\ast\sim10^{8}-10^{10}$\,G), ECME originates at radii $r\sim10^{9}-10^{10}$\,cm, where $B_{\rm loc}\sim10^{2}-10^{3}$\,G. Outside the maser growth region, the magnetospheric geometry is expected to give rise to an escape layer extending over a comparable radial scale, with a characteristic thickness $D\sim10^{7}-10^{10}$\,cm (order-of-magnitude estimate only). These path lengths correspond to electron columns sufficient to produce Faraday conversion depths $\tau_C\sim0.1-10$ at the observed frequencies, allowing intrinsically circular ECME to emerge as predominantly linear radiation. Modest source-to-source and epoch-to-epoch variations in ISM density, relative velocity, or magnetospheric loading can therefore shift $\tau_C$ between weak-conversion (circular-dominated) and strong-conversion (linear-dominated) regimes, and in extreme cases suppress detectable emission altogether, naturally accounting for both the diversity and the temporal variability of polarisation properties observed in LPRTs.

This slab should not be identified with the diffuse ISM along the line of sight, since this cannot produce significant Faraday conversion at GHz frequencies. The Galactic magnetic field strength is only a few $\mu$G, and the conversion coefficient scales as $\rho_W\propto n_e B^2$. For illustration, adopting an ISM density of $1-10$\,cm$^{-3}$ and $B\simeq5\,\mu{\rm G}$ over $D\simeq5$\,kpc, Eq.~\eqref{eq:rhos_cold} gives $\tau_{C,{\rm ISM}}\sim10^{-5}-10^{-4}$ at $\nu\sim1$\,GHz, i.e., $\tau_{C,{\rm ISM}}\ll1$, demonstrating that the ISM cannot significantly modify the polarisation state of the radiation.

Quasi-periodic fine structure is also consistent with ECME phenomenology. In ASKAP\,J$1839-0756$, substructure as narrow as $100$\,ms with a quasi-period of $2.4$\,s has been reported \citep[see Fig. 4 of ][]{Lee2025_6.45hours_LPRT_Interpulses}. Qualitatively similar fine structure is a well-established signature of ECME in many different environments. Auroral kilometric radiation from the Earth also shows richly structured, rapidly varying dynamic spectra \citep{Treumann2006_ECME, Taubenschuss2023_ECME_FineStructure}. Jupiter’s decametric emission displays narrowband drifting elements \citep{Zarka1998_AuroralEmission_OuterPlanets}, while solar coherent radio bursts exhibit sub-second spikes and fine structure, interpreted as signatures of small-scale ECME in active-region magnetic loops \citep[e.g.][]{White2024ECME_SolarBursts_FineStructure}.

A related point involves the possible presence of an incoherent radio component produced by the same non-thermal electrons that drive the maser. In early-type main-sequence stars with rigidly rotating magnetospheres, such electrons are known to generate persistent, weakly polarised incoherent synchrotron or gyrosynchrotron emission, interpreted as arising from centrifugal breakout events and large-scale magnetospheric currents \citep{Leto2021_nonthermal_Planets_stars, Shultz2022_Rotation_MRPs, Owocki2022_Centrifugal_Radio_EarlyTypeStars, Das2025_Radio_MagMassiveStars, Leto2025_NT_Radio_BA_Stars}. For the seven LPRTs discussed here, however, no convincing detection of a steady, weakly polarised or unpolarised incoherent radio component has been reported to date. In all cases, the flux densities are dominated by highly polarised, burst-like emission, while off-pulse intervals are consistent with non-detections at the sensitivity limits of the observations. Existing data, therefore, place upper limits on any persistent incoherent component at levels well below the peak burst flux densities. This difference is physically expected. Unlike early-type stars, isolated neutron stars lack dense, centrifugally supported co-rotating magnetospheres, and their magnetospheric plasma supply is instead intermittent and externally regulated, for example by weak ISM accretion. Any incoherent synchrotron or gyrosynchrotron emission from mildly relativistic electrons is therefore expected to be both quite faint and short-lived. The absence of a detected incoherent component thus does not contradict the ECME interpretation, but instead highlights a fundamental difference between neutron-star magnetospheres and the centrifugal breakout-driven magnetospheres of early-type main-sequence stars.

There are also clear limitations to the way we have discussed this emission mechanism. We have assumed centred dipoles and cold-plasma propagation. Multipolar fields or warm-plasma corrections will introduce additional complexity. Time-dependent inhomogeneities or structured inflow have not been discussed explicitly, but they are likely to influence both visibility and polarisation. A further limitation is that we have implicitly adopted a loss-cone anisotropy to drive the maser, as naturally arises in converging magnetospheric fields. In planetary auroral zones, ECME is often powered by a ``horseshoe'' (or shell) distribution generated by field-aligned acceleration \citep[e.g.][]{Ergun2000_AKR_horseshoe, Treumann2006_ECME}. It remains unclear whether isolated neutron stars can develop comparable acceleration structures; addressing this would require detailed kinetic modelling that is beyond the scope of the present work. A fully self-consistent treatment would require integrating maser growth, three-dimensional simulations of the type conducted by \citet{Leto2016_AuroralRadioEmission}, and frequency-dependent ray propagation.

Despite these caveats, the overall picture holds together. Geometry, polarisation, intermittency, energetics, fine structure, and downdrift all find natural explanations within an ECME framework. This suggests that long-period neutron stars with comparatively low magnetic fields, low space velocities, and ISM accretion may form a previously unrecognised class of coherent radio emitters (and sometimes X-ray emitters), offering new insights into neutron stars' magnetospheres beyond the classical pulsar regime.

\section{Conclusions}\label{sec:conclusions}

The properties of LPRTs are explained by ECME operating in the extended magnetospheres of long-period, slowly moving, neutron stars with comparatively low surface magnetic fields. In this framework, the GHz emission originates at radii where $B_{\rm loc}\simeq10^{2}-10^{3}$\,G, and the intrinsically narrow beaming of the maser reproduces the observed burst durations, duty cycles, and inferred luminosities. Intrinsically circular ECME becomes partially, or even predominantly, linear through Faraday conversion in an overlying magnetised escape layer, accounting for the observed diversity of polarisation states, from nearly purely circular to almost entirely linear, depending on the cumulative conversion depth along the propagation path.

A key outcome of our analysis is that neutron stars are strongly favoured over MWDs as the central engines of LPRTs detectable at kiloparsec distances. Although isolated MWDs may, in principle, host ECME, their much smaller accretion power severely limits the achievable luminosity. In addition, their vastly larger magnetic-field curvature radii suppress the degree of rotational modulation. As a result, ECME from isolated MWDs is expected to be weak, quasi-steady, and detectable only within a few hundred parsecs, consistent with existing non-detections. By contrast, neutron stars naturally combine sufficient magnetospheric curvature, compact emission geometry, and higher accretion power to generate strongly modulated, detectable bursts at Galactic distances.

Weak accretion from the ISM provides a natural and sustainable power source for ECME in isolated neutron stars, provided that the Alfv\'en radius satisfies $R_A\lesssim R_{\rm co}$, so that centrifugal inhibition does not operate. In this regime, captured material can intermittently reach the stellar surface, producing modest soft X-ray emission, while the same inflow supplies non-thermal electrons to the outer magnetosphere where the maser operates. The nearly phase-aligned radio and X-ray pulses observed from ASKAP\,J$1832-0911$ offer direct observational support for this picture, demonstrating that weak ISM accretion can simultaneously power both coherent radio bursts and surface X-ray emission without invoking a binary companion.

Energetic considerations further imply that detectable LPRTs are favoured among long-period neutron stars with comparatively low magnetic fields, $B_\ast\lesssim10^{10}$\,G, moving slowly through the ISM. Fields in the range $B_\ast\sim10^{8}-10^{10}$\,G naturally satisfy the combined requirements of (i) sufficient accretion power to sustain ECME, (ii) plasma conditions favourable for maser growth, and (iii) Alfv\'en radii small enough to avoid centrifugal inhibition. Neutron stars with substantially higher magnetic fields or large space velocities are therefore disfavoured as detectable LPRTs, except possibly for unusually nearby sources or transient episodes of enhanced local density.

Within this framework, the observed long-term ``on'' and ``off'' behaviour of many LPRTs follows naturally from modest variations in the local ISM environment. Small changes in ambient density or relative velocity can shift the Alfv\'en radius, alter the magnetospheric loading, and modulate the leakage fraction of inflowing plasma, intermittently suppressing or restoring the maser emission without requiring intrinsic changes to the neutron star itself.

The ECME model yields several testable observational predictions. (i) A weak downward drift of pulse phase with decreasing observing frequency is expected for emission occurring along dipolar field lines where $B\propto r^{-3}$, producing fractional phase shifts of the order of $\lesssim1$ per cent across the $0.7-1.5$\,GHz band, consistent with existing measurements in several LPRTs. (ii) Correlated changes in the linear and circular polarisation fractions with frequency are predicted from the strong frequency dependence of the Faraday conversion coefficient, $\rho_W\propto\nu^{-3}$. (iii) Narrow duty cycles imply beaming fractions $f_b\sim0.01-0.1$, indicating that only a small fraction of favourable lines of sight intercept the ECME beam.

If the inferred beaming fractions are representative, the currently known LPRTs may constitute only the visible tip of a much larger Galactic population of long-period, weak-field neutron stars that are otherwise electromagnetically inconspicuous. 

Future wide-field, multi-frequency radio surveys, combined with sensitive and time-resolved X-ray monitoring, will therefore be crucial for identifying additional members of this population and for testing the predicted dependence of LPRT activity on neutron-star space velocity, local ISM density, and magnetospheric loading.

Taken together, these results suggest that LPRTs open a new observational window on a largely unexplored population of neutron stars, bridging the gap between classical pulsars and thermally emitting isolated neutron stars, and providing a novel probe of magnetospheric physics and neutron star interaction with the Galactic environment.

\section*{Acknowledgements}
We wish to thank Dayal Wickramasinghe for helpful discussions. We also thank the anonymous referee for a careful and constructive report that significantly improved the clarity and presentation of this work.

\section*{Data Availability}
This work has no associated observational data. The codes that generated the illustrative figures will be shared upon request.

%%%%%%%%%%%%%%%%%%%% REFERENCES %%%%%%%%%%%%%%%%%%

\bibliographystyle{mnras}
\bibliography{ECME}

\appendix

\section{ECME essentials, resonance, and geometry}\label{app:ecme_essentials}

A key ingredient that allows ECME to operate is the development of a loss-cone anisotropy in the electron velocity distribution, arising in a magnetised plasma where the magnetic field strength $B$ increases toward the stellar surface, causing electrons gyrating about $\mathbf B$ to experience a mirror force.

Throughout this Appendix, $\mathbf B$ denotes the magnetic field vector along a field line (with magnitude $B\equiv|\mathbf B|$), while $\mathbf B_{\rm loc}$ refers to the magnetic field vector evaluated at the ECME resonance layer, with $B_{\rm loc}\equiv|\mathbf B_{\rm loc}|$ as its magnitude (see Section~\ref{sec:geometry}).

Decompose the velocity $\mathbf v$ into components parallel, $v_\parallel$, and perpendicular, $v_\perp$, to $\mathbf B$, such that $v=\sqrt{v_\perp^2+v_\parallel^2}$, and introduce the Lorentz factor $\gamma=(1-v^2/c^2)^{-1/2}$. 

The first adiabatic invariant, the magnetic moment $\mu$, is conserved, provided the magnetic field varies on spatial scales large compared to the electron gyroradius:
\begin{equation}\label{eq:1stadconst}
\mu = \frac{p_\perp^2}{2m_e B}=\text{constant},\qquad p_\perp=\gamma m_e v_\perp.
\end{equation}
In the absence of any large, quasi-static field‐aligned potential drops that generate a parallel electric field, the total particle energy is conserved, so $\gamma$ is locally constant. Since $\mu$ is constant, an increase in $B$ forces $v_\perp^2$ to increase, and thus $v_\parallel^2$ must decrease. The parallel equation of motion along the field line (coordinate $s$) is:
\begin{equation}
\frac{dp_\parallel}{dt}  =  -\,\mu\,\frac{dB}{ds},
\end{equation}
where $p_\parallel=\gamma m_e v_\parallel$. The right-hand side of the above equation defines the mirror force
\begin{equation}\label{eq:MirrorForce}
F_\parallel  =  -\,\mu\,\frac{dB}{ds},
\end{equation}
which acts opposite to the direction of increasing $B$, slowing electrons moving into stronger fields.

The pitch angle $\alpha$ is the angle between the electron's velocity $\mathbf v$ and the local magnetic field $\mathbf B$:
\begin{equation}\label{eq:pitch}
\tan\alpha=\frac{v_\perp}{|v_\parallel|},\qquad 0\le\alpha\le \frac{\pi}{2}.
\end{equation}
In a converging magnetic field such as a dipole, electrons either mirror or precipitate depending on $\alpha$. Large $\alpha$ favour magnetic trapping, while small $\alpha$ lead to loss through precipitation into the stellar atmosphere. The so-called ``loss cone'' in velocity space consists of those pitch angles that do not mirror. Conservation of the first adiabatic invariant (Eq.~\ref{eq:1stadconst}) gives its boundary:
\[
\sin^2\alpha_{\rm L}=\frac{B}{B_{\rm m}},\qquad (B_{\rm m}>B),
\]
where  $B$ is the magnetic-field strength at the particle’s position along the field line and $B_{\rm m}$ is the field at the mirror point. Continuous removal of small-$\alpha$ electrons and their replenishment by injection (Section~\ref{sec:injection}) produces the pitch-angle population inversion that drives maser growth.

Waves grow when they satisfy the cyclotron resonance with the gyrating electrons. We adopt $e>0$ for the elementary charge, so that the electron gyrofrequency evaluated at the emission site is
\begin{equation}\label{eq:gyrofrequency}
\Omega_e=\frac{eB_{\rm loc}}{m_e c},
\end{equation}
which is strictly positive. The cyclotron resonance condition \citep[e.g.][]{Stix1992_waves_in_plasmas} is
\begin{equation}\label{eq:resonance_A}
\omega - k_\parallel v_\parallel - \frac{n\,\Omega_e}{\gamma} = 0,
\end{equation}
where $n=1,2,\dots$ is the cyclotron harmonic number. Here $k_\parallel$ is the component of the wavevector along the local magnetic field $\mathbf B_{\rm loc}$ and $v_\parallel$ the electron’s parallel velocity. The product $k_\parallel v_\parallel$ represents the Doppler shift of the wave frequency in the electron’s guiding-centre frame.

For ECME, maximum growth occurs for quasi-transverse propagation ($k_\parallel \simeq0$), so the Doppler term becomes negligible and the resonance simplifies to
\begin{equation}\label{eq:resonance_short}
\omega \simeq \frac{n\,\Omega_e}{\gamma}.
\end{equation}
Equivalently, in frequency units,
\begin{equation}\label{eq:resonance_nu_obs}
\nu \simeq \left(\frac{n}{\gamma}\right)\,\nu_{ce},\qquad
\nu_{ce} = \frac{\Omega_e}{2\pi} \simeq 2.80\,\mathrm{MHz}\left(\frac{B_{\rm loc}}{1\,\mathrm{G}}\right).
\end{equation}
Hence, the observing frequency directly fixes the local magnetic field strength:
\begin{equation}\label{eq:Bloc_A}
 B_{\rm loc}[\mathrm{G}]\simeq357\,\frac{\gamma}{n}\,\nu\,[\mathrm{GHz}].
\end{equation}
We identify the wave frequency $\nu=\omega/2\pi$ with the observing frequency $\nu_{\rm obs}$ at the telescope. Gravitational redshift corrections are negligible at the relevant altitudes.

The propagation angle $\theta_{Bk}$ between the wavevector $\mathbf{k}$ (with magnitude $k\equiv|\mathbf{k}|=2\pi/\lambda$ for a monochromatic plane wave) and the local magnetic field $\mathbf{B}_{\rm loc}$ is
\begin{equation}\label{eq:costhetaBk}
\cos\theta_{Bk}
=\frac{\mathbf{k}}{|\mathbf{k}|}\!\cdot\frac{\mathbf{B}_{\rm loc}}{|\mathbf{B}_{\rm loc}|}
=\frac{k_\parallel}{k}.
\end{equation}
Thus, the quasi-transverse condition corresponds to $k_\parallel\simeq0$ ($\theta_{Bk}\simeq90^{\circ}$), and the quasi-parallel condition to $k_\perp\simeq0$ ($\theta_{Bk}\simeq0^{\circ}$).

Two effects favour quasi-transverse escape. Firstly, the ECME growth rate peaks when the cyclotron resonance ellipse in $(v_\perp,v_\parallel)$ space (Eq.~\ref{eq:resonance_A}) intersects the loss-cone boundary near $v_\parallel\simeq0$, selecting $k_\parallel\simeq0$. Secondly, in a strongly magnetised, low-density plasma, only waves travelling nearly perpendicular to $\mathbf{B}_{\rm loc}$ escape most efficiently. Together, these conditions confine the escaping radiation to hollow cones whose axes are aligned with the local $\mathbf{B}_{\rm loc}$ at the emission site and whose opening angles satisfy $\theta_{Bk}\simeq 90^{\circ}$.

The key parameter controlling both growth and escape is the ratio of the plasma frequency $\omega_p$ to gyrofrequency,
\begin{equation}\label{eq:plasma_freq_def}
\omega_{\rm p}=\left(\frac{4\pi n_e e^2}{m_e}\right)^{1/2}, \qquad
\frac{\omega_{\rm p}}{\Omega_{e}}\lesssim0.1-0.3,
\end{equation}
for which the emission can propagate quasi-transversely without encountering cut-offs or evanescence \citep{MelroseDulk1982_ECM_Sun,HewittMelrose1982_LossCone_ECM}. In this regime, the birefringent coupling coefficients ($\rho_R,\rho_W$) that mix the Stokes parameters may be modest locally, but their cumulative effect along the escape path can still produce significant Faraday conversion between $U$ and $V$.

Hence both the growth maximum and the transparency window select emission angles close to $90^{\circ}$,
\begin{equation}\label{eq:theta_cone}
\theta_{Bk}=\theta_{\rm cone}\simeq90^{\circ}-\delta\theta,
\end{equation}
producing radiation confined to a thin hollow cone whose axis is parallel to $\mathbf{B}_{\rm loc}$ (Eq.~\ref{eq:Bloc_A}) at the emission site. The small offset $\delta\theta$ (a few degrees) depends on $\omega_{\rm p}/\Omega_{e}$,  the harmonic number $n$, and the characteristic electron energy. The cone’s wall thickness $\sigma$ is the angular width over which the ECME growth rate remains appreciable and thus scales, to order of magnitude, with the fractional resonance bandwidth,
\begin{equation}
\sigma \propto \frac{\Delta\nu}{\nu},
\end{equation}
because a finite $\Delta\nu$ corresponds to a finite range of angles that satisfy the cyclotron resonance condition.

\bsp

\label{lastpage}
\end{document}